\documentclass[english,prd,superscriptaddress,nofootinbib,preprintnumbers,eqsecnum]{revtex4}

\usepackage{graphicx}
\usepackage{bm}
\usepackage{amsmath}
\usepackage{color}
\usepackage{amsfonts}

\begin{document}
\preprint{YITP-16-42}
\newcommand{\newc}{\newcommand}

\newcommand{\ben}{\begin{eqnarray}}
\newcommand{\een}{\end{eqnarray}}
\newcommand{\na}{\nabla}
\newcommand{\si}{\sigma}
\newc{\be}{\begin{equation}}
\newc{\ee}{\end{equation}}
\newc{\ba}{\begin{eqnarray}}
\newc{\ea}{\end{eqnarray}}
\newc{\bea}{\begin{eqnarray*}}
\newc{\eea}{\end{eqnarray*}}
\newc{\ie}{{\it i.e.} }
\newc{\eg}{{\it e.g.} }
\newc{\etc}{{\it etc.} }
\newc{\etal}{{\it et al.}}
\newcommand{\nn}{\nonumber}
\newc{\ra}{\rightarrow}
\newc{\lra}{\leftrightarrow}
\newc{\lsim}{\buildrel{<}\over{\sim}}
\newc{\gsim}{\buildrel{>}\over{\sim}}
\newc{\C}{{\cal C}}
\newc{\D}{{\cal D}}
\newc{\Xp}{X_{\phi}}
\newc{\Xc}{X_{\chi}}
\newc{\Mpl}{M_{\rm pl}}

\title{Screening fifth forces in generalized Proca theories}

\author{Antonio De Felice}
\affiliation{Yukawa Institute for Theoretical Physics, Kyoto University, 606-8502, Kyoto, Japan}

\author{Lavinia Heisenberg}
\affiliation{Institute for Theoretical Studies, ETH Zurich, Clausiusstrasse 47, 8092 Zurich, Switzerland}

\author{Ryotaro Kase}
\affiliation{Department of Physics, Faculty of Science, Tokyo University of Science, 1-3, Kagurazaka,
Shinjuku-ku, Tokyo 162-8601, Japan}

\author{Shinji Tsujikawa}
\affiliation{Department of Physics, Faculty of Science, Tokyo University of Science, 1-3, Kagurazaka,
Shinjuku-ku, Tokyo 162-8601, Japan}

\author{Ying-li Zhang}
\affiliation{National Astronomy Observatories, Chinese Academy of Science,
Beijing 100012, People's Republic of China}
\affiliation{Institute of Cosmology and Gravitation,
 University of Portsmouth, Portsmouth PO1 3FX, UK}

\author{Gong-Bo Zhao}
\affiliation{National Astronomy Observatories, Chinese Academy of Science,
Beijing 100012, People's Republic of China}
\affiliation{Institute of Cosmology and Gravitation,
 University of Portsmouth, Portsmouth PO1 3FX, UK}

\date{\today}

\begin{abstract}

For a massive vector field with derivative self-interactions,
the breaking of the gauge invariance allows the propagation
of a longitudinal mode in addition to the two transverse modes.
We consider generalized Proca theories with second-order
equations of motion in a curved space-time and study how
the longitudinal scalar mode of the vector field gravitates on
a spherically symmetric background. We show explicitly that
cubic-order self-interactions lead to the suppression of the
longitudinal mode through the Vainshtein mechanism. Provided
that the dimensionless coupling of the interaction is not negligible,
this screening mechanism is sufficiently efficient to give rise to tiny
corrections to gravitational potentials consistent with solar-system
tests of gravity. We also study the quartic interactions with the
presence of non-minimal derivative coupling with the Ricci scalar
and find the existence of solutions where the longitudinal mode
completely vanishes. Finally, we discuss the case in which the
effect of the quartic interactions dominates over the cubic one and
show that local gravity constraints can be satisfied
under a mild bound on the parameters of the theory.

\end{abstract}


\maketitle

\section{Introduction}

The construction of theories beyond General Relativity (GR) is
motivated not only by the ultraviolet completion of gravity but also
by the accumulating observational evidence of the late-time
cosmic acceleration.
If we modify gravity from GR, however, additional degrees of freedom
(DOF) generally arise \cite{CST,Sil,Tsuji10,Pedro,Joyce}.
To keep the theories healthy, these new DOF
should give rise to neither ghosts nor instabilities.
If the equations of motion are of second order, the lack of
higher-order derivatives forbids the propagation of
further dangerous DOF associated with Ostrogradski
instabilities \cite{Ostro}.
In the presence of one scalar degree of freedom, it is known
that Horndeski theories \cite{Horndeski} are the most general
scalar-tensor theories with second-order equations of motion
in curved space-times.
Independently of the original work, the same action was rederived by
extending the so-called Galileon action (``scalar Galileons'') \cite{Nicolis,Defa1}
to curved space-time with the second-order
property maintained \cite{Defa2,KYY,Gao,GS,CdRLH,Char,LHRKKY}.

In 1976, Horndeski derived the most general action of
an Abelian vector field with a non-minimal coupling yielding
second-order equations of motion, under the assumption that
the Maxwell equations are recovered in the flat space-time \cite{Horndeski2}.
The cosmology and the stability of such Horndeski
vector-tensor theories were recently studied
in Refs.~\cite{Barrow,Jimenez}.
There have been attempts for constructing theories of
Abelian vector fields analogous to scalar
Galileons \cite{Deser,TKK,Fleury}.
If we try to preserve the $U(1)$ gauge invariance for
one vector field and stick to second-order equations of motion,
there exists a no-go theorem stating that the Maxwell kinetic term
is the only allowed interaction \cite{Mukohyama,Mukohyama2}.
However, dropping the $U(1)$ gauge invariance allows us
to generate non-trivial terms associated with ``vector Galileons'' \cite{Heisenberg,Tasinato1}
(see also Refs.~\cite{Tasinato2,Hull,Li,Hull2,Peter,JBJTSK} for related works).

In relativistic field theory, it is well known that introduction of the mass
term for a Maxwell vector field breaks the $U(1)$ gauge invariance.
In this massive vector Proca theory, there is one propagating degree
of freedom in the longitudinal direction besides two DOF corresponding
to the transverse polarizations.
In the presence of derivative interactions like those appearing for Galileons,
it is natural to ask whether they do not modify the number of DOF in
Proca theory. In Ref.~\cite{Heisenberg}, one of the authors
derived a generalized Proca action for a vector field $A^{\mu}$
with second-order equations of motion on curved space-times.
The analysis based on the Hessian matrix showed that only three DOF
propagate as in the original Proca theory \cite{Heisenberg,Peter}.
The action has non-minimal derivative couplings to the Ricci scalar $R$ and
the Einstein tensor $G_{\mu \nu}$, whose structure is
similar to that in scalar Horndeski
theories. In fact, taking the limit $A^{\mu} \to \nabla^{\mu}\pi$,
the resulting action for the scalar field $\pi$ reproduces that
of scalar Galileons with suitable choices of free functions \cite{Heisenberg,Tasinato1}.

It was shown in Refs.~\cite{Tasinato1,Tasinato2} that
a sub-class of these generalized Proca theories can lead to the
self-acceleration of the Universe. If we apply these theories
to the present cosmic acceleration, not only a viable cosmic
expansion history could be realized but also the gravitational
interaction similar to GR could be recovered inside
the solar system. In this paper, the issue of how the vector
field gravitates in the presence of derivative
self-interactions is addressed on the spherically symmetric
space-time with a matter source.
We first show that the transverse components of
the spatial vector $A^i$ vanish on the spherically symmetric
background by imposing their regularities at the origin.
Hence the longitudinal scalar component is the only
relevant contribution to $A^i$ in addition to the time
component of $A^{\mu}$.

We study how the longitudinal propagation affects
the behavior of gravitational potentials
in the presence of the vector Galileon interactions.
We shall consider two cases: (i) the self-interacting
Lagrangian ${\cal L}_3=\beta_3X\nabla_{\mu}A^{\mu}$ exists,
and (ii) the non-minimal derivative coupling $\beta_4X^2 R$
is taken into account in the Lagrangian ${\cal L}_4$
in addition to ${\cal L}_3$.
We show that, due to derivative self-interactions,
the screening mechanism of the longitudinal mode
can be at work. This leads to the suppression of the propagation
of the fifth force in such a way that the theories
are consistent with local gravity constraints.
This is analogous to the Vainshtein mechanism \cite{Vainshtein}
for scalar Galileons \cite{Nicolis,Burrage,DKT,KKY,Kase13},
but the property of screened solutions exhibits some difference
due to the non-trivial coupling between the longitudinal mode
and the time component of $A^{\mu}$.

This paper is organized as follows.
In Sec.~\ref{HPsec} we present the action of the generalized Proca theories
in the presence of a matter source and derive the equations of motion
up to the Lagrangian ${\cal L}_4$ on general curved backgrounds.
In Sec.~\ref{spsec} we obtain the equations of motion on the spherically
symmetric background (with coefficients given in the Appendix).
In Sec.~\ref{L3sec} we derive the vector field profiles
in the presence of the Lagrangian ${\cal L}_3$ both analytically
and numerically and compute corrections to leading-order
gravitational potentials induced by the longitudinal scalar.
In Sec.~\ref{L4sec} we study the cases in which the contribution of
the Lagrangian ${\cal L}_4$ dominates over that of ${\cal L}_3$ and
also obtain analytic field profiles as well as gravitational potentials.
Sec.~\ref{consec} is devoted to conclusions.

\section{Generalized Proca theories}
\label{HPsec}

We begin with the generalized Proca theories described by
the four-dimensional action
\be
S=\int d^4x \sqrt{-g} \left( {\cal L}
+{\cal L}_m \right)\,,\qquad
{\cal L}={\cal L}_F+\sum_{i=2}^{5} {\cal L}_i\,,
\label{Lag}
\ee
where $g$ denotes the determinant of the metric tensor
$g_{\mu \nu}$, ${\cal L}_m$ the matter Lagrangian,
and ${\cal L}_F = -(1/4)F_{\mu \nu}F^{\mu \nu}$ is the
standard kinetic term of the vector field $A_{\mu}$
with $F_{\mu \nu}=\nabla_{\mu}A_{\nu}-\nabla_{\nu}
A_{\mu}$ ($\nabla_{\mu}$ is the covariant derivative operator).
The Lagrangians ${\cal L}_i$
encode the non-trivial derivative interactions \cite{Heisenberg}
\ba
 {\cal L}_2 &=& G_2(X)\,,
\label{L2}\\
{\cal L}_3 &=& G_3(X) \nabla_{\mu}A^{\mu}\,,
\label{L3}\\
{\cal L}_4 &=&
G_4(X)R+
G_{4,X}(X) \left[ (\nabla_{\mu} A^{\mu})^2
+c_2 \nabla_{\rho}A_{\sigma} \nabla^{\rho}A^{\sigma}
-(1+c_2) \nabla_{\rho}A_{\sigma}
\nabla^{\sigma}A^{\rho} \right]\,,\label{L4} \\
{\cal L}_5 &=&
G_{5}(X) G_{\mu \nu} \nabla^{\mu} A^{\nu}
-\frac16 G_{5,X}(X) [ (\nabla_{\mu} A^{\mu})^3
-3d_2 \nabla_{\mu} A^{\mu}
\nabla_{\rho}A_{\sigma} \nabla^{\rho}A^{\sigma}
-3(1-d_2) \nabla_{\mu} A^{\mu}
\nabla_{\rho}A_{\sigma} \nabla^{\sigma}A^{\rho}
\nonumber \\
& &
+(2-3d_2) \nabla_{\rho}A_{\sigma} \nabla^{\gamma}
A^{\rho} \nabla^{\sigma}A_{\gamma}
+3d_2 \nabla_{\rho}A_{\sigma} \nabla^{\gamma}
A^{\rho} \nabla_{\gamma}A^{\sigma}]\,,
\label{L5}
\ea
where
$R$ is the Ricci scalar, $G_{\mu \nu}$ is the
Einstein tensor, $G_{2,3,4,5}$ as well as $c_2,d_2$
are arbitrary functions of
\be
X \equiv -\frac12 A_{\mu} A^{\mu}\,,
\label{Xdef}
\ee
and $G_{i,X} \equiv \partial G_{i}/\partial X$. Note that we could have allowed
any contractions of the vector field $A_\mu$ with $F_{\mu\nu}$
and $F_{\mu\nu}^*$ (with $F^*$ being the dual of $F$)
in the function $G_2$, for instance in the form of
$A_\mu A_\nu F^{\mu \rho}F_\rho^\nu$...etc, or contractions
between the vector field and the Einstein tensor
$G_{\mu\nu}A^\mu A^\nu$, since they do not contain any time derivative
applying on the temporal component of the vector field,
but for the purpose of our present analysis of screened solutions
we shall simply assume  $G_2(X)$.

The Lagrangians ${\cal L}_{2,3,4,5}$ given above
keep the equations of motion up to second-order.
They can be constructed from the
Lagrangian \cite{Heisenberg,Tasinato1}
\be
\tilde{{\cal L}}_{i+2}=-\frac{1}{(4-i)!}
G_{i+2}(X) {\cal E}_{\alpha_1 \cdots \alpha_i
\gamma_{i+1 \cdots 4}}  {\cal E}^{\beta_1 \cdots \beta_i
\gamma_{i+1 \cdots 4}} \nabla_{\beta_1} A^{\alpha_1}
\cdots \nabla_{\beta_i} A^{\alpha_i}\,,\label{LG}
\ee
where $i=0,1,2,3$, and
${\cal E}_{\mu_1 \mu_2 \mu_3 \mu_4}$ is
the anti-symmetric Levi-Civita tensor.
For $i=0$ and $i=1$, we have that
${\cal L}_2=\tilde{\cal L}_2$ and
${\cal L}_3=\tilde{\cal L}_3$, respectively.
For $i=2,3$, besides the terms $\tilde{{\cal L}}_{4}$ and
$\tilde{{\cal L}}_{5}$, there are other Lagrangians
$\bar{{\cal L}}_{4}$ and $\bar{{\cal L}}_{5}$, respectively,
derived by exchanging the indices in Eq.~(\ref{LG}),
e.g., $-(1/2)F_4(X){\cal E}_{\alpha_1 \alpha_2\gamma_3\gamma_4}
{\cal E}^{\beta_1 \beta_2\gamma_3\gamma_4}
\nabla_{\beta_1}A_{\beta_2}\nabla^{\alpha_1}A^{\alpha_2}$ for $i=2$ and
$-F_5(X){\cal E}_{\alpha_1 \alpha_2\alpha_3\gamma_4}
{\cal E}^{\beta_1 \beta_2\beta_3\gamma_4}
\nabla_{\beta_1} A^{\alpha_1}
\nabla_{\beta_2} A^{\alpha_2} \nabla^{\alpha_3} A_{\beta_3}$
for $i=3$, where $F_4(X)$ and
$F_5(X)$ are arbitrary functions of $X$.
Since ${\cal L}_4=\tilde{{\cal L}}_{4}+\bar{{\cal L}}_{4}$
and ${\cal L}_5=\tilde{{\cal L}}_{5}+\bar{{\cal L}}_{5}$,
the coefficients $c_2$ and $d_2$ appearing in
Eqs.~(\ref{L4}) and (\ref{L5}) correspond to
$c_2=F_4(X)/G_4(X)$ and $d_2=F_5(X)/G_5(X)$,
respectively. Throughout this paper, we assume that
$c_2$ and $d_2$ are constants. In Eqs.~(\ref{L4}) and (\ref{L5})
the non-minimal coupling terms $G_4(X)R$ and
$G_5(X)G_{\mu \nu}\nabla^{\mu}A^{\nu}$ are
included to guarantee that the equations of motion are
of second order \cite{Heisenberg}.

The Proca Lagrangian corresponds to the functions
$G_2=m^2X$ and $G_{3,4,5}=0$,
where $m$ corresponds to the mass of the vector field.
The generalized Proca theories given by Eq.~(\ref{Lag})
generally break the $U(1)$ gauge symmetry.
It is possible to restore the gauge symmetry by introducing
a Stueckelberg field $\pi$ \cite{Stu}, as $A_{\mu} \to
A_{\mu}+\partial_{\mu} \pi$.
To zero-th order in $A_{\mu}$, we can extract
the longitudinal mode of the vector field \cite{Heisenberg}.
For the functional choices $G_2=X, G_3=X$ and
$G_4=X^2, G_5=X^2$, this procedure
gives rise to the scalar covariant Galileon Lagrangian
originally derived in Refs.~\cite{Nicolis,Defa1} by imposing
the Galilean symmetry
$\partial_{\mu} \pi \to \partial_{\mu} \pi+b_{\mu}$
in flat space-time.
The dependence on the parameters $c_2$ and $d_2$
present in Eqs.~(\ref{L4}) and (\ref{L5}) disappears
for the Stueckelberg field $\pi$.
In fact, the terms multiplied by the coefficients
$c_2$ and $d_2$ are proportional to
$G_{4,X}F_{\mu \nu}F^{\mu \nu}$ and
$G_{5,X}[(\nabla_{\lambda}A^{\lambda})
F_{\mu \nu}F^{\mu \nu}/2+\nabla_{\mu}A_{\nu}
\nabla^{\nu}A_{\rho}F^{\rho \mu}]$,
respectively, which are both expressed in terms
of $F_{\mu \nu}$ \cite{Heisenberg,Peter}.

In the following we focus on theories given by the action
(\ref{Lag}) up to the Lagrangian ${\cal L}_4$.
We do not consider the Lagrangian ${\cal L}_5$
due to its complexity, but we leave such an analysis for
a future work.
We define the energy-momentum tensor of the matter
Lagrangian ${\cal L}_m$, as
\be
T_{\mu \nu}^{(m)}=-\frac{2}{\sqrt{-g}}
\frac{\delta(\sqrt{-g}{\cal L}_m)}
{\delta g^{\mu \nu}}\,.
\ee
Assuming that matter is minimally coupled to gravity,
there is the continuity equation
\be
\nabla^{\mu}T_{\mu \nu}^{(m)}=0\,.
\label{Tcon}
\ee
Variation of the action (\ref{Lag})
with respect to $g^{\mu \nu}$ and $A_{\nu}$
leads to
\be
\delta S=\int d^4 x \sqrt{-g}
\left[ \left( {\cal G}_{\mu \nu}-\frac12 T_{\mu \nu}^{(m)}
\right) \delta g^{\mu \nu}
+{\cal A}^{\nu} \delta A_{\nu} \right]\,,
\ee
where
\be
{\cal G}_{\mu \nu} \equiv
\frac{\delta {\cal L}}{\delta g^{\mu \nu}}
-\frac12 g_{\mu \nu}{\cal L}\,,
\qquad
{\cal A}^{\nu}
\equiv \frac{\delta {\cal L}}{\delta A_{\nu}}\,.
\ee

The equation of motion of the gravity sector on
general curved space-times is given by
\be
{\cal G}_{\mu \nu}=\frac12 T_{\mu \nu}^{(m)}\,,
\label{Ein}
\ee
with
\be
{\cal G}_{\mu \nu}=
{\cal G}_{\mu \nu}^{(F)}
+\sum_{i=2}^{4}{\cal G}_{\mu \nu}^{(i)}\,.
\ee
Here each term comes from the standard kinetic term and the Lagrangians
(\ref{L2})-(\ref{L4}), as
\ba
{\cal G}_{\mu \nu}^{(F)}
&=&\frac14 g_{\mu \nu} \left( \na_{\rho} A_{\si}
\na^{\rho} A^{\si}-\na_{\rho} A_{\si}
\na^{\si} A^{\rho} \right)
-\frac12 \left[ \na_{\rho}A_{\mu} \na^{\rho}A_{\nu}
+\na_{\mu}A_{\rho} \na_{\nu}A^{\rho}
-2\na_{\rho}A_{(\nu} \na_{\mu)}A^{\rho}
\right] \,,\label{GF}\\
{\cal G}_{\mu \nu}^{(2)}
&=&-\frac12 g_{\mu \nu}G_2-\frac12 G_{2,X}
A_{\mu}A_{\nu}\,,\\
{\cal G}_{\mu \nu}^{(3)}
&=&-\frac12 G_{3,X}
\left[ A_{\mu}A_{\nu} \na_{\rho}A^{\rho}
+g_{\mu \nu}A^{\lambda}A_{\rho}
\nabla_{\lambda}A^{\rho}
-2A_{\rho}A_{(\mu}\na_{\nu)}A^{\rho} \right]\,,\\
{\cal G}_{\mu \nu}^{(4)}
&=& G_4 G_{\mu \nu}-\frac12 G_{4,X}A_{\mu}A_{\nu}R
\nonumber \\
& &+\frac12 G_{4,X} g_{\mu \nu}[(\na_{\rho}A^{\rho})^2
-(2+c_2)\na_{\rho}A_{\si} \na^{\rho}A^{\si}
+(1+c_2)\na_{\rho}A_{\si} \na^{\si}A^{\rho}
-2A_{\rho} \square A^{\rho}
+2A^{\rho}\na_{\rho}\na_{\si}A^{\si}]
\nonumber \\
& &+G_{4,X}[(1+c_2)\na_{\mu}A_{\rho}\na_{\nu}A^{\rho}
-\na_{\rho}A^{\rho} \na_{(\mu}A_{\nu)}
-(1+2c_2)\na_{\rho}A_{(\nu}\na_{\mu)}A^{\rho}
+(1+c_2)\na_{\rho}A_{\mu}\na^{\rho}A_{\nu} \nonumber \\
& &+A_{\rho}\na_{(\mu}\na_{\nu)}A^{\rho}
-A^{\rho}\na_{\rho}\na_{(\mu}A_{\nu)}
+A_{(\nu}\square A_{\mu)}
-2A_{(\nu}\na_{\mu)}\na_{\si}A^{\si}
+A_{(\mu}\nabla_{\rho}\nabla_{\nu)}A^{\rho}] \nonumber \\
&& -\frac12 G_{4,XX} \{ A_{\mu}A_{\nu}
[(\na_{\rho}A^{\rho})^2+c_2\na_{\rho}A_{\si}
\na^{\rho}A^{\si}-(1+c_2)\na_{\rho}A_\si \na^{\si}A^{\rho}]
+2A_{\rho}A_{\si}\na_{\mu}A^{\rho}\na_{\nu}A^{\si}
\nonumber \\
&&-2A_{\alpha}\na_{\rho}A^{\alpha}
[A^{\rho}\na_{(\mu}A_{\nu)}-A_{(\nu}\na_{\mu)}A^{\rho}-A_{(\nu}\na^{\rho}A_{\mu)}
-2g_{\mu \nu}A^{[\rho}\na^{\si]}A_{\si}]
-4A_{\alpha}(\na_{\si}A^{\si})
A_{(\nu}\na_{\mu)}A^{\alpha} \}\,,\label{G4}
\ea
where $\nabla_{(\mu} A_{\nu)} \equiv
(\nabla_{\mu}A_{\nu}+\nabla_{\nu}A_{\mu})/2$ and
$A^{[\rho}\nabla^{\sigma]}A_{\sigma} \equiv
(A^{\rho}\nabla^{\sigma}A_{\sigma}-
A^{\sigma}\nabla^{\rho}A_{\sigma})/2$.
The equation of motion for the vector field $A_{\nu}$
corresponds to ${\cal A}^{\nu}=0$, i.e.,
\ba
& &
\nabla_{\mu}F^{\mu \nu}-G_{2,X}A^{\nu}
+2G_{3,X} A^{[\mu}\nabla^{\nu]}A_{\mu}
-RG_{4,X}A^{\nu}-2G_{4,X} \left[ \nabla^{\nu}
\nabla_{\mu} A^{\mu}+c_2 \square A^{\nu}
-(1+c_2)\nabla^{\mu} \nabla^{\nu}A_{\mu} \right]
\nonumber \\
& & -G_{4,XX}[ A^{\nu} \left\{ (\nabla_{\mu}A^{\mu})^2
+c_2 \nabla_{\rho}A_{\sigma} \nabla^{\rho}A^{\sigma}
-(1+c_2)\nabla_{\rho}A_{\sigma}\nabla^{\sigma}A^{\rho}
\right\} \nonumber \\
& &
-2A_{\rho}\nabla^{\nu}A^{\rho} \nabla_{\mu}A^{\mu}
-2c_2 A_{\rho} \nabla^{\mu}A^{\rho}\nabla_{\mu}A^{\nu}
+2(1+c_2)A_{\rho} \nabla^{\mu}A^{\rho} \nabla^{\nu}
A_{\mu}]=0\,.
\label{eqA}
\ea

In GR we have $G_4=M_{\rm pl}^2/2$, where
$M_{\rm pl}$ is the reduced Planck mass, so
${\cal G}_{\mu \nu}^{(4)}$ simply reduces to
$(M_{\rm pl}^2/2)G_{\mu \nu}$.
Existence of the vector field with derivative self-couplings
induces additional gravitational interactions with matter  through Eq.~(\ref{Ein}).
We shall study whether such a fifth force can be suppressed
in local regions with a matter source.

\section{Equations of motion on the spherically symmetric
background}
\label{spsec}

We derive the equations of motion on the
spherically symmetric and static background
described by the line element
\be
ds^{2}=-e^{2\Psi(r)}dt^{2}+e^{2\Phi(r)}dr^{2}
+r^{2} \left( d\theta^{2}+\sin^{2}\theta\,
d\varphi^{2} \right)\,,
\label{line}
\ee
where $\Psi(r)$ and $\Phi(r)$ are the gravitational potentials
that depend on radius $r$ from the center of sphere.
For the matter Lagrangian ${\cal L}_m$,  we consider
the perfect fluid with the energy-momentum tensor
$T^{\mu}_{\nu}={\rm diag}(-\rho_m,P_m,P_m,P_m)$,
where $\rho_m$ is the energy density and $P_m$ is
the pressure.
Then, the matter continuity equation (\ref{Tcon}) reads
\be
P_m'+\Psi' (\rho_m+P_m)=0\,,
\label{Pmeq}
\ee
where a prime represents a derivative
with respect to $r$.

We write the vector field $A^{\mu}$ in the form
\be
A^{\mu}=\left( \phi, A^{i}\right)\,,
\ee
where $i=1,2,3$. {}From Helmholtz's theorem,
we can decompose the spatial components
$A^{i}$ into the transverse and longitudinal modes, as
\be
A_{i}=A_{i}^{(T)}+\nabla_{i}\chi \,,
\ee
where $A_{i}^{(T)}$ obeys the traceless condition
$\nabla^i A_{i}^{(T)}=0$ and $\chi$ is the longitudinal scalar.
On the spherically symmetric configuration, it is required that
the $\theta$ and $\varphi$ components of $A_i^{(T)}$
(i.e., $A_2^{(T)}$ and $A_3^{(T)}$) vanish.
Then, the traceless condition gives the following relation
\be
{A_1^{(T)}}'+\frac{2}{r}A_1^{(T)}-\Phi' A_1^{(T)}=0\,,
\ee
whose solution is given by
\be
A_1^{(T)}=C\,\frac{e^{\Phi}}{r^2}\,,
\ee
where $C$ is an integration constant.
For the regularity of $A_1^{(T)}$ at $r=0$, we require that $C=0$.
This discussion shows that the transverse vector $A_{i}^{(T)}$
vanishes, so we only need to focus on the
propagation of the longitudinal mode, i.e., $A_i=\nabla_i \chi$.
Then, the components of $A^{\mu}$
on the spherical coordinate ($t,r,\theta,\varphi$) are given by
\be
A^{\mu}=\left(\phi(r), e^{-2\Phi}\chi'(r),0,0 \right)\,.
\label{Amu}
\ee

The $(0,0)$, $(1,1)$ and $(2,2)$
components of Eq.~(\ref{Ein}) reduce,
respectively, to\footnote{We note that the $(0,1)$ component
of Eq.~(\ref{Ein}) reduces to the same form as Eq.~(\ref{F1}).}
\ba
&&
\C_1 \Psi'^2+\left(\C_2+\frac{\C_3}{r}\right) \Psi'
+\left(\C_4+\frac{\C_5}{r}\right) \Phi'
+\C_6 +\frac{\C_7}{r}+\frac{\C_8}{r^2}=
-e^{2\Phi}\rho_m\,,
\label{eom00}\\
&&
\C_9 \Psi'^2+\left(\C_{10}+\frac{\C_{11}}{r}\right) \Psi'
+\C_{12}+\frac{\C_{13}}{r}+\frac{\C_{14}}{r^2}=
e^{2\Phi}P_m\,,
\label{eom11}\\
&&
\C_{15} \Psi''+\C_{16} \Psi'^2+\C_{17} \Psi' \Phi'
+\left(\C_{18}+\frac{\C_3/4+\C_{15}}{r}\right) \Psi'
+\left(-\frac{\C_{13}}{2}+\frac{\C_{19}}{r}\right) \Phi'
+\C_{20}+\frac{\C_{21}}{r}\notag\\
&&=e^{2\Phi}P_m\,,
\label{eom22}
\ea
where the coefficients $\C_i$ ($i=1,2, \cdots, 21$)
are given in the Appendix.
The mass term (\ref{Xdef}) can be decomposed as
$X=\Xp+\Xc$, where
\be
\Xp\equiv\frac{1}{2}e^{2\Psi}\phi^2\,,\qquad
\Xc\equiv-\frac{1}{2}e^{-2\Phi}\chi'^2\,.
\ee

The $\nu=0$ and $\nu=1$ components of Eq.~(\ref{eqA})
reduce, respectively, to
\ba
&&\D_1(\Psi''+\Psi'^2)+\D_2 \Psi'\Phi'+\left(\D_3+\frac{\D_4}{r}\right) \Psi'
+\left(\D_5+\frac{\D_6}{r}\right) \Phi'+\D_7+\frac{\D_8}{r}+\frac{\D_9}{r^2}=0\,,
\label{F0}\\
&&
\D_{10}\Psi'^2+\left(\D_{11}+\frac{\D_{12}}{r}\right) \Psi'
+\D_{13}+\frac{\D_{14}}{r}+\frac{\D_{15}}{r^2}=0\,,
\label{F1}
\ea
where we introduced the short-cut notations for convenience
\ba
&&
\D_1=2\phi(2c_2G_{4,X}-1)\,,\quad
\D_2=2\phi[1-2c_2(G_{4,X}+2\Xc G_{4,XX})]\,,
\notag\\
&&
\D_3=\phi\chi' G_{3,X}-\phi'[3-2c_2(3G_{4,X}+2\Xp G_{4,XX})]
-4c_2 e^{-2\Phi} \phi\chi'\chi''G_{4,XX}\,,
\notag\\
&&
\D_4=4\phi(2c_2G_{4,X}-2\Xc G_{4,XX}-1)\,,\quad
\D_5=-\phi\chi' G_{3,X}+\phi'[1-2 c_2 (G_{4,X}+2\Xc G_{4,XX})]\,,
\notag\\
&&
\D_6=4\phi(G_{4,X}+2\Xc G_{4,XX})\,,
\notag\\
&&
\D_7=e^{2\Phi}\phi G_{2,X}+\phi\chi''G_{3,X}
-\phi''(1-2c_2G_{4,X})
+c_2(e^{2\Psi}\phi\phi'^2-2e^{-2\Phi}\phi'\chi'\chi'')G_{4,XX}\,,
\notag\\
&&
\D_8=2\phi\chi'G_{3,X}-2\phi'(1-2c_2G_{4,X})
+4e^{-2\Phi}\phi\chi'\chi''G_{4,XX}\,,\quad
\D_9=-2\phi[(1-e^{2\Phi})G_{4,X}+2\Xc G_{4,XX}]\,,
\notag\\
&&
\D_{10}=8c_2e^{-2\Phi}\chi'\Xp G_{4,XX}\,,\quad
\D_{11}=2(\Xc-\Xp)G_{3,X}+4c_2e^{2\Psi-2\Phi}\phi\phi'\chi'G_{4,XX}\,,
\notag\\
&&
\D_{12}=4e^{-2\Phi}\chi'[G_{4,X}+2(\Xc-\Xp)G_{4,XX}]\,,\quad
\D_{13}=-\chi'G_{2,X}-e^{2\Psi}\phi\phi'G_{3,X}
+c_2e^{2\Psi-2\Phi}\phi'^2\chi'G_{4,XX}\,,
\notag\\
&&
\D_{14}=4\Xc G_{3,X}-4e^{2\Psi-2\Phi}\phi\phi'\chi'G_{4,XX}\,,\quad
\D_{15}=-2\chi'[(1-e^{-2\Phi})G_{4,X}-2e^{-2\Phi}\Xc G_{4,XX}]\,.
\ea
Among the six equations of motion (\ref{Pmeq}),
(\ref{eom00})-(\ref{eom22}), and (\ref{F0})-(\ref{F1}),
five of them are independent.
For a given density profile $\rho_m$ of matter, solving five
independent equations of motion leads to the solutions to
$P_m,\Psi,\Phi,\phi,\chi$ with appropriate boundary conditions.

For the consistency with local gravity experiments within the
solar system, we require that the gravitational potentials
$\Psi$ and $\Phi$ need to be close to those in GR.
In GR without the vector field $A^{\mu}$, we have
$G_2=G_3=0$, $G_4=M_{\rm pl}^2/2$ and
$\phi=0=\chi'$, so Eqs.~(\ref{eom00}) and (\ref{eom11}) read
\ba
&&
\frac{2M_{\rm pl}^2}{r}\Phi_{\rm GR}'
-\frac{M_{\rm pl}^2}{r^2}
\left( 1-e^{2\Phi_{\rm GR}} \right)
=e^{2\Phi_{\rm GR}} \rho_m\,,
\label{GR1}\\
&&
\frac{2M_{\rm pl}^2}{r}\Psi_{\rm GR}'
+\frac{M_{\rm pl}^2}{r^2}
\left( 1-e^{2\Phi_{\rm GR}} \right)
=e^{2\Phi_{\rm GR}} P_m\,.
\label{GR2}
\ea
Since $\Phi_{\rm GR}$ and $\Psi_{\rm GR}$ would be the leading-order
contributions to gravitational potentials under the operation of the
screening mechanism, we first derive their solutions
inside and outside a compact body.
We assume that the change of $\rho_m$ occurs rapidly
at the distance $r_*$, so that the matter density can be
approximated as $\rho_m(r) \simeq \rho_0$ for $r<r_*$
and $\rho_m(r) \simeq 0$ for $r>r_*$.
This configuration is equivalent to that of the
Schwarzschild interior and exterior solutions.
For $r<r_*$, integration of Eq.~(\ref{Pmeq}) leads to
$P_m=-\rho_m+{\cal C}e^{-\Psi(r)}$, where ${\cal C}$
is an integration constant known by imposing the condition $P_m(r_*)=0$.

Matching the interior and exterior solutions of $\Psi$ and
$\Phi$ at $r=r_*$ with appropriate boundary conditions (at
$r=0$ and $r \to \infty$), the gravitational potentials
inside and outside the body are given by
\be
e^{\Psi_{\rm GR}}=\frac32 \sqrt{1-\frac{\rho_0 r_*^2}
{3M_{\rm pl}^2}}
-\frac12 \sqrt{1-\frac{\rho_0 r^2}{3M_{\rm pl}^2}}\,,\qquad
e^{\Phi_{\rm GR}}= \left( 1-\frac{\rho_0 r^2}{3M_{\rm pl}^2}\right)^{-1/2}\,,
\label{Sin}
\ee
for $r<r_*$, and
\be
e^{\Psi_{\rm GR}}=\left( 1 -\frac{\rho_0 r_*^3}
{3M_{\rm pl}^2 r} \right)^{1/2}\,,
\qquad
e^{\Phi_{\rm GR}}=\left( 1 -\frac{\rho_0 r_*^3}
{3M_{\rm pl}^2 r} \right)^{-1/2}\,,
\label{Sout}
\ee
for $r>r_*$. In the following, we employ the weak
gravity approximation under which $|\Psi|$ and $|\Phi|$ are much smaller than 1, i.e.,
\be
\Phi_* \equiv
\frac{\rho_0 r_*^2}{M_{\rm pl}^2} \ll 1\,.
\label{weakcon}
\ee
This condition means that the Schwarzschild radius
of the source $r_g \approx \rho_0 r_*^3/M_{\rm pl}^2$
is much smaller than $r_*$.
Then, the solutions (\ref{Sin}) and (\ref{Sout}) reduce,
respectively, to
\be
\Psi_{\rm GR} \simeq
\frac{\rho_0}{12M_{\rm pl}^2} \left( r^2-3r_*^2 \right)\,,\qquad
\Phi_{\rm GR} \simeq
\frac{\rho_0 r^2}{6M_{\rm pl}^2}\,,
\label{Sin2}
\ee
for $r<r_*$, and
\be
\Psi_{\rm GR} \simeq
-\frac{\rho_0r_*^3}{6M_{\rm pl}^2r}
\,,\qquad
\Phi_{\rm GR} \simeq
\frac{\rho_0r_*^3}{6M_{\rm pl}^2r}\,,
\label{Sout2}
\ee
for $r>r_*$.
For the theories with the action (\ref{Lag}), the vector field
interacts with gravity through the derivative terms $\Psi'', \Psi', \Phi'$ in Eqs.~(\ref{F0}) and (\ref{F1}).
The leading-order contributions of such gravitational interactions follow from the derivatives
$\Psi_{\rm GR}'', \Psi_{\rm GR}', \Phi_{\rm GR}'$
of the GR solutions (\ref{Sin})-(\ref{Sout}).
Then, we can integrate
Eqs.~(\ref{F0}) and (\ref{F1}) to obtain the solutions
to $\phi$ and $\chi'$. The next-to-leading order corrections
to $\Psi$ and $\Phi$ can be derived by substituting
the solutions of $\phi$ and $\chi'$ into Eqs.~(\ref{eom00})
and (\ref{eom11}).
In Secs.~\ref{L3sec} and \ref{L4sec} we apply this
procedure to concrete theories.

\section{Theories with the cubic Lagrangian}
\label{L3sec}

Let us first consider theories in which the function
$G_4$ corresponds only to the Einstein-Hilbert term, i.e.,
\be
G_4=\frac{M_{\rm pl}^2}{2}\,,
\ee
where $M_{\rm pl}$ is the reduced Planck mass.
In this case the $G_{4,X}$ term in the Lagrangian
${\cal L}_4$ vanishes, but the Lagrangian
${\cal L}_3$ gives rise to a non-trivial gravitational
interaction with the vector field.
The equations of motion (\ref{F0}) and (\ref{F1})
reduce, respectively to
\ba
& &
\frac{1}{r^2} \frac{d}{dr} (r^2 \phi')-e^{2\Phi}G_{2,X} \phi
-G_{3,X}\phi\, \frac{1}{r^2} \frac{d}{dr} (r^2 \chi') \nonumber \\
& &
+2\phi \left( \Psi''+\Psi'^2-\Psi'\Phi' \right)
-\left( \phi \chi' G_{3,X}-3\phi'-\frac{4\phi}{r} \right)
\Psi'
+\left( \phi \chi' G_{3,X}-\phi' \right)\Phi'=0\,,
\label{F0eq} \\
& &
\chi' G_{2,X}+\left( e^{2\Psi} \phi \phi'+\frac{2}{r}
e^{-2\Phi} \chi'^2 \right) G_{3,X}
+\left( e^{2\Psi} \phi^2+e^{-2\Phi}\chi'^2 \right)G_{3,X}\Psi'=0\,.
\label{F1eq}
\ea
For concreteness, we shall focus on the theories
given by the functions
\be
G_2(X)=m^2 X\,,\qquad
G_3(X)=\beta_3 X\,,
\ee
where $m$ is the mass of the vector field,
and $\beta_3$ is a dimensionless constant.
The choice of $G_3(X)$ given above is related with
that of scalar Galileons.
In what follows, we obtain analytic solutions to
Eqs.~(\ref{F0eq}) and (\ref{F1eq}) under the
approximation of weak gravity.

\subsection{Analytic vector field profiles}
\subsubsection{Solutions for $r<r_*$}

For the distance $r$ smaller than $r_*$, we substitute
the derivatives of Eq.~(\ref{Sin2}) into Eqs.~(\ref{F0eq})
and (\ref{F1eq}) to derive leading-order solutions
to $\phi$ and $\chi'$.
The terms containing $e^{2\Psi}$ and $e^{-2\Phi}$
provide the contributions linear in $\Psi$ and $\Phi$
[say, $\Psi \phi \phi' G_{3,X}$ in Eq.~(\ref{F1eq})].
After deriving analytic
solutions to $\phi$ and $\chi'$, however, we can show
that such terms give rise to contributions much smaller than the
leading-order solutions. Hence it is consistent to employ the
approximations $e^{2\Psi} \simeq 1$ and
$e^{-2\Phi} \simeq 1$ in Eqs.~(\ref{F0eq})
and (\ref{F1eq}), such that
\ba
& &
\frac{d}{dr} (r^2 \phi')-m^2 r^2 \phi
-\beta_3\phi\, \frac{d}{dr} (r^2 \chi')
+\frac{\rho_0}{6M_{\rm pl}^2} \left[
6\phi+r \left( \phi'+\beta_3 \chi' \phi \right)
\right] r^2 \simeq 0\,,
\label{F0eq2}\\
& &
m^2 \chi'+\beta_3\left( \phi \phi'+\frac{2}{r}\chi'^2+
\frac{\rho_0 \phi^2}{6M_{\rm pl}^2}r \right)
\simeq 0\,.
\label{F2eq2}
\ea

From Eq.~(\ref{F2eq2}) it follows that
\be
\chi'=\frac{m^2r}{4\beta_3} \left[ -1+
\sqrt{1-\frac{8\beta_3^2}{m^4 r} \left( \phi\phi'
+\frac{\rho_0 \phi^2}{6M_{\rm pl}^2}r
\right)} \right]\,.
\label{chige}
\ee
The sign of (\ref{chige}) has been chosen in such a way that
$\chi'$ vanishes for $\beta_3/m^2 \to 0$, which can
be regarded as the GR limit.
Since we are interested in how the screening mechanism
is at work in the presence of the Lagrangian ${\cal L}_3$
for a very light field (e.g., the vector field
associated with the late-time cosmic acceleration), we take
another limit $\beta_3/m^2 \to \infty$ in the discussion below.
In other words, we focus on the case $m \to 0$ with
a non-zero dimensionless coupling $\beta_3$.
For $\beta_3>0$, Eq.~(\ref{chige}) reduces to
\be
\chi'=\sqrt{-\frac{r}{2} \left( \phi\phi'
+\frac{\rho_0\phi^2}{6M_{\rm pl}^2}r \right)}\,.
\label{chiso}
\ee
For the consistency of Eq.~(\ref{chiso}) we require
the condition $\phi \phi'<0$.

We search for solutions where the scalar potential $\phi$
does not vary much with respect to $r$, i.e.,
\be
\phi(r)=\phi_0+f(r)\,,\qquad
|f(r)| \ll |\phi_0|\,,
\label{phiap}
\ee
where $\phi_0$ is a constant and $f(r)$ is a function of $r$.
We also focus on the case where $\phi(r)$ decreases
with the growth of $r$, such that $\phi'(r)<0$
with $\phi_0>0$.
In Eq.~(\ref{F0eq2}) we also neglect the terms
$r(\phi'+\beta_3 \chi' \phi)$ relative to $6\phi$.
The validity of this approximation can be checked
after deriving the solutions to $\phi$ and $\chi'$.
Substituting Eq.~(\ref{chiso}) into Eq.~(\ref{F0eq2})
with Eq.~(\ref{phiap}), we obtain the integrated solution
\be
r^2f'-\beta_3 \phi_0^{3/2} r^2
\sqrt{-\frac{r}{2} \left( f'+\frac{\rho_0 \phi_0 r}
{6M_{\rm pl}^2} \right)}
+\frac{\rho_0\phi_0}{3M_{\rm pl}^2}r^3=C\,,
\ee
where $C$ is a constant.
Under the boundary condition $\phi'(0)=0$, we can
fix $C=0$ and hence
\be
f'-\beta_3 \phi_0^{3/2}
\sqrt{-\frac{r}{2} \left( f'+\frac{\rho_0 \phi_0 r}
{6M_{\rm pl}^2} \right)}
=-\frac{\rho_0\phi_0}{3M_{\rm pl}^2}r\,.
\label{fso}
\ee
Clearly, there is a solution of the form
$f'(r) \propto -r$. Substituting the solution
$f(r)=-Br^2$ into Eq.~(\ref{fso}), we find that the positive
constant $B$, which remains finite in the limit
$\beta_3 \to \infty$, is given by
\be
B=\frac{\rho_0 \phi_0}{6M_{\rm pl}^2}
{\cal F}(s_{\beta_3})\,,
\ee
where
\ba
s_{\beta_3} &\equiv&
\frac{3(\beta_3 \phi_0M_{\rm pl})^2}{4\rho_0}\,,
\label{sbeta}\\
{\cal F}(s_{\beta_3}) &\equiv& (1+s_{\beta_3})
\left( 1-\sqrt{\frac{s_{\beta_3}}{1+s_{\beta_3}}} \right)\,.
\ea
Then, we obtain the following analytic field profiles
\ba
\phi(r)
&=&\phi_0 \left[ 1
-{\cal F}(s_{\beta_3})\frac{\rho_0}{6M_{\rm pl}^2}
r^2 \right]\,,\label{phiinL3} \\
\chi'(r)
&=&\sqrt{\frac{\rho_0\phi_0^2}{6M_{\rm pl}^2}
\left[ {\cal F}(s_{\beta_3})-\frac12\right]}\,r\,.
\label{chiinL3}
\ea

As $s_{\beta_3}$ increases from 0 to $\infty$, the function
${\cal F}(s_{\beta_3})$ decreases from 1 to $1/2$.
This means that the terms inside the square root of
Eq.~(\ref{chiso}) remains positive.
Since ${\cal F}(s_{\beta_3})\rho_0 r^2/
(6M_{\rm pl}^2) \ll 1$ from the condition
(\ref{weakcon}) of weak gravity, the solution (\ref{phiinL3})
is consistent with the assumption (\ref{phiap}).
In the limit that $s_{\beta_3} \ll 1$, the field profiles
(\ref{phiinL3}) and (\ref{chiinL3}) reduce,
respectively, to
\be
\phi(r) \simeq \phi_0 \left( 1-
\frac{\rho_0}{6M_{\rm pl}^2}r^2 \right)\,,
\qquad
\chi'(r) \simeq \sqrt{\frac{\rho_0\phi_0^2}
{12M_{\rm pl}^2}}r\,,
\label{inss}
\ee
whereas, for $s_{\beta_3} \gg 1$, it follows that
\be
\phi(r) \simeq \phi_0 \left( 1-
\frac{\rho_0}{12M_{\rm pl}^2}r^2 \right)\,,
\qquad
\chi'(r) \simeq \frac{\rho_0}{6\beta_3 M_{\rm pl}^2}r\,.
\label{inls}
\ee
The amplitude of $\chi'(r)$ in Eq.~(\ref{inls}) is about
$s_{\beta_3}^{-1/2}$ times smaller than
that in Eq.~(\ref{inss}).
For a larger coupling $|\beta_3|$, the screening effect
is efficient to suppress the propagation of the longitudinal
mode. On using the solutions (\ref{phiinL3}) and (\ref{chiinL3}),
we can confirm that the terms
$r(\phi'+\beta_3 \chi' \phi)$ in Eq.~(\ref{F0eq2}) is
much smaller than $6\phi$ and that the approximations
$e^{2\Psi} \simeq 1$ and $e^{-2\Phi} \simeq 1$
employed in Eq.~(\ref{F2eq2}) are also justified.

\subsubsection{Solutions for $r>r_*$}

Employing the GR solution (\ref{Sout2}) of gravitational potentials in the regime $r>r_*$ and
substituting them into Eqs.~(\ref{F0eq}) and (\ref{F1eq}),
it follows that
\ba
& &
\frac{d}{dr} (r^2 \phi')-m^2 r^2 \phi
-\beta_3\phi\, \frac{d}{dr} (r^2 \chi')
+\frac{\rho_0r_*^3}{9M_{\rm pl}^4r^2} \left[
\rho_0 r_*^3 \phi+3M_{\rm pl}^2 r^2
(2\phi'-\beta_3\chi' \phi)
\right]  \simeq 0\,,
\label{F0eq3}\\
& &
m^2 \chi'+\beta_3\left( \phi \phi'+\frac{2}{r}\chi'^2+
\frac{\rho_0 \phi^2r_*^3}{6M_{\rm pl}^2r^2} \right)
\simeq 0\,.
\label{F2eq3}
\ea
Taking the $m \to 0$ limit and considering the branch
$\chi'>0$, Eq.~(\ref{F2eq3}) gives the following relation
\be
\chi'=\sqrt{-\frac{r}{2} \left( \phi\phi'
+\frac{\rho_0\phi^2r_*^3}{6M_{\rm pl}^2r^2} \right)}\,.
\label{chiso2}
\ee
The term $(\rho_0r_*^3)^2\phi/(9M_{\rm pl}^4r^2)$
in Eq.~(\ref{F0eq3}) is at most $\Phi_*$
times as small as the term $\rho_0\phi/M_{\rm pl}^2$
in Eq.~(\ref{F0eq2}). Moreover, after deriving the solutions
to $\phi$ and $\chi'$,
we can confirm that the contributions $3M_{\rm pl}^2 r^2
(2\phi'-\beta_3\chi' \phi)$ in Eq.~(\ref{F0eq3}) is at most
of the order of $\rho_0 r_*^3 \phi$.
Hence it is a good approximation to neglect the terms
inside the square bracket of Eq.~(\ref{F0eq3}).
Substituting Eq.~(\ref{chiso2}) into Eq.~(\ref{F0eq3})
with the approximation (\ref{phiap}) and matching
the integrated solution at $r=r_*$ on account of Eq.~(\ref{fso}), we obtain
\be
r^2\phi'-\beta_3 \phi_0^{3/2}r^2
\sqrt{-\frac{r}{2} \left( \phi'+\frac{\rho_0\phi_0r_*^3}
{6M_{\rm pl}^2 r^2} \right)}
\simeq -\frac{\rho_0 \phi_0r_*^3}{3M_{\rm pl}^2}\,.
\ee

More explicitly, the field derivative $\phi'$ can be
expressed as
\be
\phi'(r)=-\frac{\rho_0\phi_0r_*^3}{3M_{\rm pl}^2r^2}
{\cal F}(\xi)\,,\qquad \xi \equiv s_{\beta_3} \frac{r^3}{r_*^3}\,.
\label{xidef}
\ee
{}From Eq.~(\ref{chiso2}) the longitudinal mode is given by
\be
\chi'(r)=\sqrt{\frac{\rho_0r_*^3\phi_0^2}{6M_{\rm pl}^2r}
\left[ {\cal F}(\xi)-\frac12 \right]}\,.
\label{chid}
\ee

If $s_{\beta_3} \gg 1$, then $\xi \gg 1$ for $r>r_*$.
In this case it follows that
\be
\phi'(r) \simeq -\frac{\rho_0\phi_0r_*^3}{6M_{\rm pl}^2r^2}\,,
\qquad \chi'(r) \simeq \frac{\rho_0r_*^3}
{6\beta_3 M_{\rm pl}^2 r^2}\,.
\label{slarge}
\ee

If $s_{\beta} \lesssim 1$, there is the transition
radius $r_V$ at which the $r$ dependence of the longitudinal
mode changes. The radius $r_V$ can be identified by the
condition $\xi=1$, i.e.,
\be
r_V=\frac{r_*}{s_{\beta_3}^{1/3}}\,.
\label{rV}
\ee
For the distance $r_*<r \ll r_V$ we have ${\cal F} \simeq 1$,
so the solutions reduce to
\be
\phi'(r) \simeq -\frac{\rho_0\phi_0r_*^3}{3M_{\rm pl}^2r^2}\,,\qquad
\chi'(r)\simeq
\sqrt
{\frac{\rho_0r_*^3\phi_0^2}{12M_{\rm pl}^2r}}\,.
\label{ssmall}
\ee
For $r \gg r_V$ we have $\xi \gg 1$ and hence the resulting
solutions are given by Eq.~(\ref{slarge}).
In this regime, the longitudinal mode $\chi'(r)$ decreases faster than
that  for $r_*<r \ll r_V$ with a suppressed amplitude.
The distance $r_V$ can be regarded as the Vainshtein radius above
which $\chi'(r)$ starts to decay quickly.
If $|\beta_3|$ obeys the condition $s_{\beta_3} \gg 1$,
$\chi'(r)$ is strongly suppressed both inside and outside
the body due to the Vainshtein mechanism, see
Eqs.~(\ref{inls}) and (\ref{slarge}).
Meanwhile, for $s_{\beta_3} \lesssim 1$, the screening
of the longitudinal mode manifests for the distance $r>r_V$.
The fact that the suppression of the longitudinal mode
occurs {\it outside} the radius
$r_V$ for small $|\beta_3|$ is a unique feature of
vector Galileons.

\subsection{Numerical solutions for the vector field}

To confirm the validity of the analytic solutions derived above,
we shall numerically solve Eqs.~(\ref{F0eq}) and (\ref{F1eq})
coupled with the gravitational Eqs.~(\ref{eom00})-(\ref{eom22}).
For concreteness we consider the density
distribution given by
\be
\rho_m(r)=\rho_0 e^{-ar^2/r_*^2}\,,
\label{denprofile}
\ee
where $a$ is a positive constant of the order of 1.
With this profile, the matter density starts to decrease
significantly for $r \gtrsim r_*$.
For the numerical purpose, it is convenient to
introduce the following dimensionless quantities
\be
x=\frac{r}{r_*}\,,\qquad
y=\frac{\phi}{\phi_0}\,,\qquad
z=\frac{\chi'}{\phi_0}\,,
\ee
where $\phi_0$ is the value of $\phi$
at $r=0$. In the massless limit with $G_3=\beta_3X$,
we can express Eqs.~(\ref{F0eq})
and (\ref{F1eq}) in the forms
\ba
& &
\frac{d^2y}{dx^2}+\frac{2}{x}\frac{dy}{dx}
-\beta_3 r_*\phi_0 y\left( \frac{dz}{dx} +\frac{2}{x}z
\right)+2y \left[ \frac{d^2\Psi}{dx^2}+\left(
\frac{d\Psi}{dx} \right)^2-\frac{d\Psi}{dx}
\frac{d\Phi}{dx} \right] \nonumber \\
& &
-\left( \beta_3 r_*\phi_0 yz-3\frac{dy}{dx}
-\frac{4}{x} y \right)\frac{d\Psi}{dx}
+\left( \beta_3 r_*\phi_0 yz-\frac{dy}{dx}
\right) \frac{d\Phi}{dx}=0\,, \label{nso1}\\
& &
z=e^{\Psi+\Phi} \sqrt{-xy \left( \frac{dy}{dx}
+y\frac{d\Psi}{dx} \right) \left( 2+x \frac{d\Psi}{dx}
\right)^{-1}}\,,
\label{nso2}
\ea
where the quantity $\beta_3 r_*\phi_0$ is related with
$s_{\beta_3}$ as $\beta_3 r_*\phi_0=
\sqrt{4s_{\beta_3}\Phi_*/3}$.
We take the $x$ derivative of Eq.~(\ref{nso2})
and then eliminate the term $dz/dx$ by combining it
with Eq.~(\ref{nso1}) to obtain the second-order
equation for $y(x)$. To derive the leading-order
gravitational potentials $\Phi_{\rm GR}$ and
$\Psi_{\rm GR}$, we also solve Eqs.~(\ref{GR1})
and (\ref{GR2}) with a vanishing pressure $P_m$.
This procedure gives rise to the solutions derived
under the weak gravity approximation, e.g., Eq.~(\ref{Sin}).
Numerically, we confirmed that the approximation
substituting $\Phi_{\rm GR}$ and $\Psi_{\rm GR}$
into Eqs.~(\ref{nso1}) and (\ref{nso2}) provides
practically identical results to those
obtained by solving full Eqs.~(\ref{eom00})-(\ref{eom22}).

\begin{figure}
\begin{center}
\includegraphics[width=3.5in]{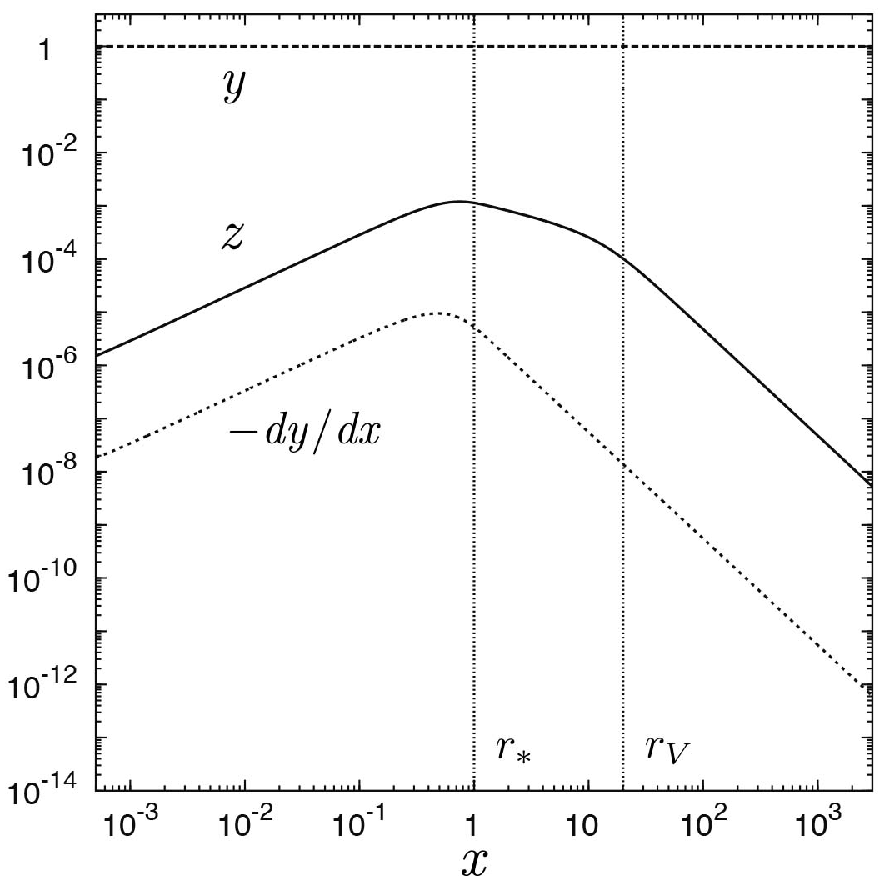}
\includegraphics[width=3.5in]{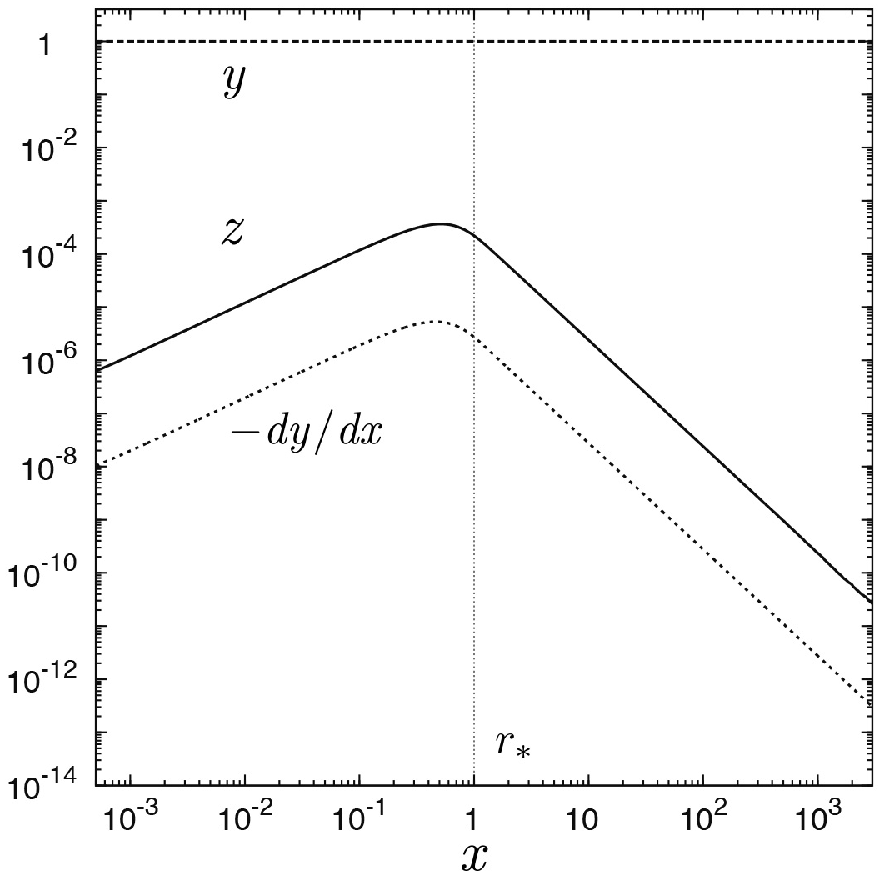}
\caption{The numerical solutions to $y=\phi/\phi_0$,
$-dy/dx$, and $z=\chi'/\phi_0$ as a function of $x=r/r_*$
for the matter profile $\rho_m=\rho_0 e^{-4r^2/r_*^2}$
with $\Phi_*=10^{-4}$.
Each panel corresponds to $s_{\beta_3}=10^{-4}$ (left) and
$s_{\beta_3}=1$ (right), respectively.
The boundary conditions of $\Psi$, $\Phi$, $y$, and
$dy/dx$ are chosen to be consistent with Eqs.~(\ref{Sin}) and
(\ref{phiinL3}) at $x=10^{-3}$. The vertical
lines represent the scales $r=r_*$ and $r_V=20r_*$ (left panel)
and the scale $r=r_*$ (right panel).
\label{fig1}}
\end{center}
\end{figure}

In Fig.~\ref{fig1} we plot the field profile for
$\rho_m=\rho_0 e^{-4r^2/r_*^2}$ and $\Phi_*=10^{-4}$
with two different values of $s_{\beta_3}$.
The boundary conditions of $y$ and $dy/dx$ around
the center of body are chosen to match
with Eq.~(\ref{phiinL3}).
As we see in Fig.~\ref{fig1}, both
$-\phi'(r)$ and $\chi'(r)$ linearly grow in $r$
for the distance smaller than $r_*$.
The left panel of Fig.~\ref{fig1} corresponds to
$s_{\beta_3}=10^{-4}$, in which case
the solutions to $\phi(r)$ and $\chi'(r)$ are
well described by Eq.~(\ref{inss}) in the regime
$r<r_*$. For $s_{\beta_3}$ larger than the order of 1,
the longitudinal mode $\chi'(r)$ tends to be suppressed
according to Eq.~(\ref{inls}). The right panel of
Fig.~\ref{fig1}, which corresponds to $s_{\beta_3}=1$,
is the case in which the suppression of $\chi'(r)$ occurs
in a mild way for $r<r_*$.

For the distance $r$ larger than $r_*$, both
$-\phi'(r)$ and $\chi'(r)$ start to decrease
with the growth of $r$.
When $s_{\beta_3}=10^{-4}$, the distance $r_V$
is of the order of $10\,r_*$.
Hence the solutions to $\phi'(r)$ and
$\chi'(r)$ are given by Eq.~(\ref{ssmall}) for
$r_*<r \lesssim 10\,r_*$
and by Eq.~(\ref{slarge}) for $r \gtrsim 10\,r_*$.
In the left panel of Fig.~\ref{fig1}, we can confirm that
the qualitative behavior of $\chi'(r)$ changes around
$r \approx 10\,r_*$
(i.e., from $\chi'(r) \propto r^{-1/2}$ to
$\chi'(r) \propto r^{-2}$).
Note that $|\phi'(r)|$ decreases as
$|\phi'(r)| \propto r^{-2}$ for $r>r_*$.

When $s_{\beta_3}=1$, as seen in the right panel of
Fig.~\ref{fig1}, we find that there is almost no intermediate regime
corresponding to the solution $\chi'(r) \propto r^{-1/2}$ and that
the longitudinal mode decreases as $\chi'(r) \propto r^{-2}$ for $r>r_*$.
This reflects the fact that, even when $s_{\beta_3}=O(1)$,
the quantity $\xi$ in Eq.~(\ref{xidef})
quickly becomes much larger than 1 with the growth
of $r~(>r_*)$.
Then, for $s_{\beta_3} \gtrsim 1$, the solutions
in the regime $r>r_*$ are well approximated by
Eq.~(\ref{slarge}).
For increasing $|\beta_3|$, the suppression for the amplitude
of $\chi'(r)$ tends to be more significant outside the body.

In Fig.~\ref{fig1} we also find that $\phi(r)$ stays nearly
constant in the whole regime of interest.
This is associated with the fact that the $r$-dependent
correction to $\phi(r)$ is at most of the order of
$\phi_0 \Phi_*$, i.e., much smaller than $\phi_0$
under the weak gravity approximation.
The numerical solutions to $\phi(r)$ and $\chi'(r)$ are
fully consistent with the analytic field profiles derived
under the assumption (\ref{phiap}), so we resort to
the analytic solutions for discussing corrections to
the leading-order gravitational potentials
in Sec.~\ref{correctionsec}.

\subsection{Corrections to gravitational potentials}
\label{correctionsec}

We compute the corrections to $\Phi_{\rm GR}$ and
$\Psi_{\rm GR}$ induced by the longitudinal propagation
of the vector field.
Since the leading-order gravitational potentials obey
Eqs.~(\ref{GR1}) and (\ref{GR2}), we can express
Eqs.~(\ref{eom00}) and (\ref{eom11}) in the forms
\ba
&&
\frac{2M_{\rm pl}^2}{r}\Phi'
-\frac{M_{\rm pl}^2}{r^2}
\left( 1-e^{2\Phi} \right)
=e^{2\Phi} \rho_m+\Delta_{\Phi}\,,
\label{GR1d}\\
&&
\frac{2M_{\rm pl}^2}{r}\Psi'
+\frac{M_{\rm pl}^2}{r^2}
\left( 1-e^{2\Phi} \right)
=e^{2\Phi} P_m+\Delta_{\Psi}\,,
\label{GR2d}
\ea
where $\Delta_{\Phi}$ and $\Delta_{\Psi}$ are
correction terms.
We are interested in the behavior of gravitational potentials outside
a compact object ($r \gtrsim r_*$), so we
employ the solutions (\ref{xidef}) and (\ref{chid})
with the leading-order potentials (\ref{Sout2}) to estimate
the corrections $\Delta_{\Phi}$ and $\Delta_{\Psi}$.
Note that $\phi(r)$ is given by Eq.~(\ref{phiap}) with
$|f(r)|$ at most of the order of $\phi_0\Phi_*$.

Let us first consider the case $s_{\beta_3} \gtrsim 1$.
Since the solutions to $\phi'(r)$ and $\chi'(r)$ are
approximately given by Eq.~(\ref{slarge}), it follows that
\be
\Delta_{\Phi} \simeq \frac{5\Phi_*^2\phi_0^2 r_*^2}
{72r^4}\,,\qquad
\Delta_{\Psi} \simeq
-\frac{\Phi_*^2 \phi_0^2 r_*^2}{72r^4}\,,
\ee
where we used the condition $\xi \gg 1$.
Integrations of Eqs.~(\ref{GR1d}) and (\ref{GR2d}) lead to
\be
\Phi(r) \simeq \frac{\Phi_*r_*}{6r}
\left[ 1 -\frac{5\Phi_*}{24}
\left( \frac{\phi_0}{M_{\rm pl}} \right)^2
\frac{r_*}{r}\right]\,,\qquad
\Psi(r) \simeq  -\frac{\Phi_*r_*}{6r}
\left[ 1 -\frac{\Phi_*}{8}
\left( \frac{\phi_0}{M_{\rm pl}} \right)^2
\frac{r_*}{r}\right]\,.
\ee
To recover the behavior close to GR in the solar
system, we require that
$\Phi_*(\phi_0/M_{\rm pl})^2 (r_*/r) \ll 1$.
Under this condition, the post-Newtonian parameter
$\gamma \equiv -\Phi/\Psi$ is given by
\be
\gamma \simeq 1-\frac{\Phi_*}{12}
\left( \frac{\phi_0}{M_{\rm pl}} \right)^2
\frac{r_*}{r}\,.
\ee
The local gravity experiments give the bound
$|\gamma-1|<2.3 \times 10^{-5}$ \cite{Will}.
For the Sun ($\Phi_* \simeq 10^{-6}$) we have
$|\gamma-1| \simeq 10^{-7} (\phi_0/M_{\rm pl})^2
(r_*/r)$, so the experimental bound is well satisfied
for $\phi_0 \lesssim M_{\rm pl}$.
We also note that the deviation of $\gamma$ from 1
decreases for larger $r$.

We proceed to the case $s_{\beta_3} \lesssim 1$.
On using the solutions (\ref{ssmall}) for the distance
$r_*<r<r_V$, the correction terms in Eqs.~(\ref{GR1d})
and (\ref{GR2d}) read
\be
\Delta_{\Phi} \simeq \frac{\sqrt{3\Phi_*r_*}\beta_3\phi_0^3}
{4r^{3/2}}\,,\qquad
\Delta_{\Psi} \simeq
\frac{\sqrt{3}(\Phi_*r_*)^{5/2}\beta_3\phi_0^3}
{432r^{7/2}}\,,
\ee
which means that $\Delta_{\Psi}$ is about $10^{-2}\Phi_*^2(r_*/r)^2$ times
as small as $\Delta_{\Phi}$. Integrating
Eqs.~(\ref{GR1d}) and (\ref{GR2d}), the gravitational potentials are given by
\be
\Phi(r) \simeq \frac{\Phi_*r_*}{6r}
\left[ 1+\sqrt{s_{\beta_3}} \left(\frac{\phi_0}
{M_{\rm pl}}\right)^2 \left( \frac{r}{r_*} \right)^{3/2}
\right]\,,\qquad
\Psi(r) \simeq -\frac{\Phi_*r_*}{6r}
\left[ 1-2\sqrt{s_{\beta_3}} \left(\frac{\phi_0}
{M_{\rm pl}}\right)^2 \left( \frac{r}{r_*} \right)^{3/2}
\right]\,,
\ee
where we used $s_{\beta_3}$ instead of $\beta_3$.
The correction to $\Psi_{\rm GR}(r)=-\Phi_*r_*/(6r)$
is negligibly small for $\phi_0 \lesssim M_{\rm pl}$.
The post-Newtonian parameter can be estimated as
\be
\gamma \simeq 1+
3\sqrt{s_{\beta_3}} \left(\frac{\phi_0}
{M_{\rm pl}}\right)^2 \left( \frac{r}{r_*} \right)^{3/2}\,,
\ee
which increases for larger $r$.
The maximum value of $|\gamma-1|$ is reached
at the distance $r=r_V$, i.e.,
$|\gamma-1|_{\rm max} \simeq 3(\phi_0/M_{\rm pl})^2$.
To satisfy the experimental bound
of $\gamma$ at this radius,
we require that
\be
\phi_0 \lesssim 3 \times 10^{-3} M_{\rm pl}\,.
\label{phi0}
\ee
For $r>r_V$ the solutions of the vector field change to
Eq.~(\ref{slarge}), so the parameter $|\gamma-1|$
starts to decrease.

The above discussion shows that, when
$s_{\beta_3} \gtrsim 1$,
the extra gravitational interaction induced by the longitudinal
mode $\chi'(r)$ is suppressed due to the Vainshtein mechanism.
If $s_{\beta_3} \lesssim 1$, the screening mechanism
of the fifth force is at work only at $r>r_V$, so the field value
$\phi_0$ is constrained as Eq.~(\ref{phi0}) for the consistency
with local gravity tests at $r=r_V$.
If $r_V$ is larger than the solar-system scale
($r_{\rm solar} \sim 10^{14}$\,cm), the upper bound
of $\phi_0$ gets weaker than Eq.~(\ref{phi0}).
For the Sun ($r_* \sim 10^{11}$\,cm), we have
$r_V>r_{\rm solar}$ for $s_{\beta_3} \lesssim 10^{-9}$.
In the limit that $s_{\beta_3} \to 0$ the distance $r_V$
goes to infinity, so there is no upper bound of $\phi_0$.

{}From Eq.~(\ref{sbeta}) the following relation holds
\be
\sqrt{s_{\beta_3}} \simeq 2.5 \times 10^{45}\,
\beta_3 \frac{\phi_0}{M_{\rm pl}}
\sqrt{\frac{1~{\rm g/cm}^3}{\rho_0}}\,.
\ee
For the Sun ($\rho_0 \approx 100~{\rm g/cm}^3$),
we have $\sqrt{s_{\beta_3}} \approx 10^{44}\beta_3
\phi_0/M_{\rm pl}$.
Even if $\phi_0$ is much smaller than the order
of $M_{\rm pl}$, it is natural to satisfy the condition $s_{\beta_3} \gtrsim 1$
except for a very tiny coupling $\beta_3$.
In this sense, we can say that the screening mechanism,
which occurs for $s_{\beta_3} \gtrsim 1$, is very generic in
the presence of a non-vanishing coupling $\beta_3$.

\section{Theories with the cubic and quartic Lagrangians}
\label{L4sec}

In this section we study the theories given by the functions
\be
G_2(X)=m^2X\,,\qquad
G_3(X)=\beta_3 X\,,\qquad
G_4(X)=\frac{M_{\rm pl}^2}{2}+\beta_4 X^2\,,
\label{L4lag}
\ee
where $m$, $\beta_3$, $\beta_4$ are constants
($\beta_4$ has a dimension of [mass]$^{-2}$).
Our interest is how the vector Galileon
term $\beta_4 X^2$ modifies the screening mechanism
discussed in the Sec.~\ref{L3sec}.
For the functions (\ref{L4lag}), the vector field equations
of motion (\ref{F0}) and (\ref{F1}) read
\ba
& &
\frac{1}{r^2} \frac{d}{dr} \left( r^2\phi' \right)
-e^{2\Phi}m^2\phi
+2\phi \left( \Psi''+\Psi'^2-\Psi'\Phi' \right)
+\left( 3\phi'+\frac{4\phi}{r} \right)\Psi'-\phi'\Phi'
-\beta_3 \phi \left[ \frac{1}{r^2}\frac{d}{dr}
\left( r^2\chi' \right)+\left( \Psi'-\Phi' \right)\chi' \right]  \nonumber \\
& &
-\frac{2\beta_4e^{-2\Phi}\phi}{r^2}
\left[ 4r\chi'\chi'' +e^{2\Psi+2\Phi}\phi^2
\left( e^{2\Phi}-1+2r\Phi' \right)-\chi'^2
\left\{ e^{2\Phi}-3+2r(3\Phi'-2\Psi') \right\} \right] \nonumber
 \\
& &
+\frac{2\beta_4c_2e^{-2\Phi}}{r} \biggl[
2r\chi'\chi'' (\phi'+2\phi\Psi')+\chi'^2 \left \{
4\phi\Psi'+\left[ \phi''+2\phi (\Psi''-3\Psi'\Phi'+\Psi'^2)
\right]r+\phi' (2-3r\Phi'+3r\Psi' ) \right\} \nonumber \\
& &
-e^{2\Psi+2\Phi}\phi \left\{ r\phi'^2+\phi\phi'
(2-r\Phi'+5r\Psi')+\phi( 4\phi \Psi'+[\phi''+
2\phi(\Psi''-\Psi'\Phi'+\Psi'^2)]r) \right\} \biggr]=0\,,
\label{F0full}\\
& &
m^2\chi'+\beta_3 \left[ e^{2\Psi} (\phi \phi'+\phi^2 \Psi')
+e^{-2\Phi}\chi'^2 \left( \frac{2}{r}+\Psi' \right)
\right] +
\frac{2\beta_4\chi'}{r} \biggl[
e^{2\Psi}\frac{\phi^2}{r} (1-e^{-2\Phi}) \nonumber \\
& &
+e^{2\Psi-2\Phi}
(4\phi\phi'-c_2r\phi'^2-4c_2\phi\phi'r\Psi'
-4c_2\phi^2r \Psi'^2+2\phi^2 \Psi')
-e^{-2\Phi} \frac{\chi'^2}{r}
(1-3e^{-2\Phi}-6r\Psi'e^{-2\Phi}) \biggr]=0\,.
\label{F1full}
\ea

If $\beta_3=0$, then there is a solution
$\chi'=0$ to Eq.~(\ref{F1full}).
This means that, in the absence of the Lagrangian
${\cal L}_3$, the $\beta_4 X^2$ term admits the solution
where the longitudinal mode completely vanishes.
In what follows, we shall consider the theories
with $\beta_3 \neq 0$ and $\beta_4 \neq 0$
by dealing with the coupling $\beta_3$
as a small correction to the solution $\chi'=0$.
As we will see below, the longitudinal mode $\chi'$
does not completely vanish in such cases.

As for the coupling $\beta_4$, the condition under which
the term $\beta_4 X^2 R$ in $G_4(X)R$ is subdominant
to the Einstein-Hilbert term $M_{\rm pl}^2R/2$ gives
\be
|\beta_4|\phi^4 \ll M_{\rm pl}^2\,.
\label{b4con0}
\ee
More specifically, we focus on the case in which
the coupling $\beta_4$ is in the range
\be
|\beta_4| \phi^2 \ll 1\,,
\label{b4con}
\ee
under which Eq.~(\ref{b4con0}) is satisfied for
$|\phi| \lesssim M_{\rm pl}$.
We also assume that the constant $|c_2|$ is
at most of the order of 1.

\subsection{Vector field profiles}

To derive analytic solutions to the vector field,
we employ the weak gravity approximation
($\Phi_* \ll 1$) and expand Eqs.~(\ref{F0full}) and
(\ref{F1full}) up to first order in $\Psi$, $\Phi$, and
their derivatives. Analogous to the discussion in
Sec.~\ref{L3sec}, we search for the solutions in the
form (\ref{phiap}) with $\chi'^2$ suppressed relative
to $\phi^2$, i.e.,
\be
|r\phi'| \ll |\phi|\,,\qquad
\chi'^2 \ll \phi^2\,.
\label{rphichi}
\ee
The consistency of these approximations can be checked
after deriving analytic solutions of $\phi$ and $\chi'$.
Under this approximation scheme, Eqs.~(\ref{F0full})
and (\ref{F1full}) reduce, respectively, to
\ba
& &
\left( 1-2c_2 \beta_4 \phi^2 \right) \frac{d}{dr}
\left( r^2 \phi' \right)-m^2 r^2\phi
-\beta_3 \phi \frac{d}{dr} (r^2 \chi')
-4\beta_4 \phi \frac{d}{dr} \left( r\chi'^2 \right) \nonumber \\
& &
+2\phi \frac{d}{dr} \left( r^2 \Psi' \right)
+\beta_3 \phi \chi' r^2 (\Phi'-\Psi')
-4\beta_4 \phi^3 \left[ \Phi+r(\Phi'+2c_2\Psi'+c_2r\Psi'')
\right] \simeq 0\,,
\label{F0L4ap1}\\
& &
\chi' \left[ \frac{4\beta_4}{r} \left\{ 2\phi \phi'+
\frac{\chi'^2}{r}+\frac{\phi^2}{r} (\Phi+r\Psi') \right\}
+m^2 \right] \simeq
-\beta_3 \left( \phi \phi'+\frac{2\chi'^2}{r}+\phi^2 \Psi'
\right)\,.
\label{F0L4ap2}
\ea
In the following, we take the massless limit $m \to 0$.
Then, Eq.~(\ref{F0L4ap2}) can be expressed as
\be
\chi'=-\frac{\beta_3r}{4\beta_4}
\frac{r\phi \phi'+2\chi'^2+\phi^2 r\Psi'}
{2r\phi \phi'+\chi'^2+\phi^2 (\Phi+ r\Psi')}\,.
\label{chiL4}
\ee
\subsubsection{In the regime $r<r_*$}

For the distance $r<r_*$ the leading-order gravitational potentials
are given by Eq.~(\ref{Sin2}), so Eq.~(\ref{chiL4}) reads
\be
\chi' = -\frac{\beta_3r}{4\beta_4}
\frac{r\phi \phi'+2\chi'^2+\rho_0\phi^2 r^2/(6M_{\rm pl}^2)}
{2r\phi \phi'+\chi'^2+\rho_0\phi^2 r^2/(3M_{\rm pl}^2)}\,.
\label{chiL4d}
\ee
If the condition
\be
\chi'^2 \ll r|\phi \phi'|
\label{chi2con}
\ee
is satisfied, Eq.~(\ref{chiL4d}) reduces to
\be
\chi'(r) \simeq -\frac{\beta_3}{8\beta_4}r\,,
\label{chiL4in}
\ee
whose magnitude linearly grows in $r$.
In the limit that $\beta_3 \to 0$, $\chi'$ vanishes
as expected.
Under the assumption (\ref{phiap}),
the field $\phi$ stays nearly a constant value $\phi_0$.
On using the solution (\ref{chiL4in}), Eq.~(\ref{F0L4ap1})
is integrated to give
\be
\left( 1-2c_2 \beta_4 \phi_0^2 \right)r^2 \phi'
+\frac{\beta_3^2 \phi_0r^3}{16\beta_4} \simeq
-\frac{\rho_0 \phi_0r^3}{3M_{\rm pl}^2}
\left[ 1-2(1+c_2)\beta_4 \phi_0^2 \right] \,.
\label{phiineq}
\ee
Provided that
\be
\delta \equiv
\frac{3\beta_3^2 M_{\rm pl}^2}{16\beta_4 \rho_0} \ll 1\,,
\label{b4con1}
\ee
the term containing $\beta_3$ in Eq.~(\ref{phiineq})
is sub-dominant relative to other terms.
Then, we obtain the following solution
\be
\phi'(r) \simeq -\frac{\rho_0\phi_0 r}{3M_{\rm pl}^2}
\frac{1-2(1+c_2)\beta_4\phi_0^2}
{1-2c_2 \beta_4 \phi_0^2}\,,
\label{philinear}
\ee
which is close to $\phi'(r) \simeq -\rho_0 \phi_0 r/
(3M_{\rm pl}^2)$. On using this solution with $\Phi_* \ll 1$,
it follows that $|r\phi'| \ll |\phi|$ for $r<r_*$.
The condition (\ref{chi2con}) translates to
\be
\varepsilon \equiv
\frac{3\beta_3^2 M_{\rm pl}^2}
{64\beta_4^2 \rho_0 \phi_0^2} \ll 1\,.
\label{b4con2}
\ee
Since $\varepsilon$ is related to $\delta$ in
Eq.~(\ref{b4con1}) as $\delta=4\varepsilon \beta_4\phi_0^2$,
the condition (\ref{b4con2}) is tighter than  (\ref{b4con1})
under the assumption (\ref{b4con}).
We also note that, under the condition (\ref{chi2con})
with $|r\phi'| \ll |\phi|$, the second relation of
Eq.~(\ref{rphichi}) is automatically satisfied.

\subsubsection{In the regime $r_*<r<r_t$}

For $r>r_*$ the leading-order gravitational potentials are
given by Eq.~(\ref{Sout2}), so integration of Eq.~(\ref{F0L4ap1}) leads to
\be
\left( 1-2c_2 \beta_4 \phi_0^2 \right)r^2 \phi'
-\beta_3 \phi_0r^2 \chi'-4\beta_4\phi_0 r \chi'^2
\simeq
-\frac{\rho_0 \phi_0r_*^3}{3M_{\rm pl}^2}\,,
\label{phiL4out}
\ee
whereas Eq.~(\ref{chiL4}) reduces to
\be
\chi' \simeq -\frac{\beta_3r}{4\beta_4}
\frac{r\phi \phi'+2\chi'^2+\rho_0\phi^2r_*^3
/(6M_{\rm pl}^2 r)}
{2r\phi \phi'+\chi'^2+\rho_0\phi^2r_*^3
/(3M_{\rm pl}^2 r)}\,.
\label{chiL4out}
\ee
Unlike Eq.~(\ref{phiineq}), the r.h.s. of
Eq.~(\ref{phiL4out}) is constant.
For the distance $r<r_*$ the field derivative $|\phi'|$
linearly grows in $r$ as Eq.~(\ref{philinear}),
but $|\phi'|$ starts to decrease
for $r>r_*$. Meanwhile, as long as the condition
(\ref{chi2con}) is satisfied, Eq.~(\ref{chiL4out})
gives the following solution
\be
\chi'(r) \simeq -\frac{\beta_3}{8\beta_4}r\,,
\label{chiinter}
\ee
so that $|\chi'|$ continues to grow.
Substituting this solution into Eq.~(\ref{phiL4out}),
we obtain
\be
\phi'(r) \simeq -\frac{\rho_0\phi_0r_*^3}
{3M_{\rm pl}^2(1-2c_2 \beta_4 \phi_0^2)r^2}
\left( 1+\delta \frac{r^3}{r_*^3} \right)\,.
\label{phiinter}
\ee
Under the condition (\ref{b4con1}) the second term in
the bracket of Eq.~(\ref{phiinter}) is much smaller than 1
around $r=r_*$.
Provided that $\delta r^3/r_*^3 \ll 1$, the leading-order
solution of Eq.~(\ref{phiinter}) decreases for larger $r$.

Substituting the approximate solutions $\phi \simeq \phi_0$ and
$\phi' \simeq -\rho_0\phi_0r_*^3/(3M_{\rm pl}^2r^2)$
into Eq.~(\ref{chiL4out}), it follows that
\be
\chi' \simeq -\frac{\beta_3 r}{2\beta_4}
\frac{\chi'^2-\rho_0\phi_0^2r_*^3/(12M_{\rm pl}^2r)}
{\chi'^2-\rho_0\phi_0^2r_*^3/(3M_{\rm pl}^2r)}\,.
\label{chisim1}
\ee
The increase of $|\chi'|$ gradually saturates as $\chi'^2$
approaches the value $\rho_0\phi_0^2r_*^3/(12M_{\rm pl}^2r)$.
We define the transition distance $r_t$ according to the
condition $\chi'^2(r_t)=\rho_0\phi_0^2r_*^3/(12M_{\rm pl}^2r_t)$.
On using the solution (\ref{chiinter}), we obtain
\be
r_t=\frac{1}{(4\varepsilon)^{1/3}}r_*\,,
\label{rt}
\ee
which is larger than $r_*$ under the condition (\ref{b4con2}).
Around $r=r_t$ the growth of $|\chi'|$ changes to decrease.

\subsubsection{In the regime $r>r_t$}

As $|\chi'|$ decreases for the distance $r>r_t$, the l.h.s.
of Eq.~(\ref{F0L4ap2}), which is multiplied by the factor
$4\beta_4\chi'/r$, becomes sub-dominant to the r.h.s.,
so that we obtain
\be
\phi\phi'+\frac{2}{r}\chi'^2+\frac{\rho_0\phi^2r_*^3}
{6M_{\rm pl}^2r^2} \simeq 0\,.
\label{chiext2}
\ee
Since the term $-4\beta_4\phi_0 r \chi'^2$ in Eq.~(\ref{phiL4out})
is negligible relative to $-\beta_3 \phi_0r^2 \chi'$, we have
\be
\left( 1-2c_2 \beta_4 \phi_0^2 \right)r^2 \phi'
-\beta_3 \phi_0r^2 \chi'
\simeq -\frac{\rho_0 \phi_0r_*^3}{3M_{\rm pl}^2}\,.
\label{phiext2}
\ee
Apart from the small difference of the coefficient
in front of $r^2 \phi'$, the system described by
Eqs.~(\ref{chiext2}) and (\ref{phiext2}) has the same
structure as that studied in Sec.~\ref{L3sec}.
Physically, this means that the effect of the Lagrangian
${\cal L}_3$ manifests itself for the distance $r>r_t$.

Following the same procedure as that in Sec.~\ref{L3sec},
we obtain the following field profiles
\ba
\phi'(r) &=& -\frac{\rho_0\phi_0r_*^3}{3M_{\rm pl}^2r^2}
{\cal G} (\eta)\,,\label{phirt} \\
\chi'(r) &=& \pm \sqrt{\frac{\rho_0\phi_0^2r_*^3}
{6M_{\rm pl}^2r}
\left[{\cal G} (\eta)-\frac12 \right]}\,,
\label{chirt}
\ea
where
\ba
\eta &=& \frac{s_{\beta_3}}{1-2c_2\beta_4 \phi_0^2}
\frac{r^3}{r_*^3}\,,\\
{\cal G}(\eta) &=&
\frac{1+\eta}{1-2c_2\beta_4 \phi_0^2}
\left[ 1-\sqrt{1-\frac{1+(1-2c_2\beta_4 \phi_0^2)\eta}
{(1+\eta)^2}} \right]\,.
\label{Geta}
\ea
In Eq.~(\ref{Geta}) we have chosen the branch where
${\cal G}(\eta)$ does not diverge in the limit that $\eta \to \infty$.
If the ratio $\beta_3/\beta_4$ is positive (negative), then
$\chi'$ has a negative (positive) sign.
On using Eq.~(\ref{rt}), the quantity $\eta$
can be expressed as
\be
\eta=\frac{4\beta_4^2\phi_0^4}{1-2c_2\beta_4 \phi_0^2}
\frac{r^3}{r_t^3}\,.
\ee
At the distance $r=r_t$ we have  that
$\eta \simeq 4\beta_4^2\phi_0^4 \ll 1$,
but $\eta$ increases for larger $r$.
The distance $r_v$ at which $\eta$ is equivalent
to 1 can be estimated as
\be
r_v \simeq
\frac{1}{(4 \beta_4^2 \phi_0^4)^{1/3}}r_t\,.
\ee
For the distance $r_t<r<r_v$, Eqs.~(\ref{phirt}) and
(\ref{chirt}) reduce, respectively, to
\ba
\phi'(r) &\simeq& -\frac{\rho_0\phi_0r_*^3}
{3M_{\rm pl}^2(1-2c_2 \beta_4 \phi_0^2)r^2} \,,\\
\chi'(r) &\simeq& \pm \sqrt{\frac{\rho_0\phi_0^2r_*^3}
{12M_{\rm pl}^2 r} \frac{1+2c_2 \beta_4 \phi_0^2}
{1-2c_2 \beta_4 \phi_0^2}}\,.
\label{chifi}
\ea
The behavior of $\phi'(r)$ is practically unchanged compared
to Eq.~(\ref{phiinter}).
The solution (\ref{chifi}) smoothly matches with Eq.~(\ref{chiinter})
at $r=r_t$ with the amplitude $|\chi'(r_t)| \simeq
\sqrt{\rho_0 \phi_0^2 r_*^3/(12M_{\rm pl}^2 r_t)}$.

For the distance $r>r_v$ we obtain the following solution
\ba
\phi'(r) &\simeq&
-\frac{\rho_0\phi_0r_*^3}{6M_{\rm pl}^2r^2}\,,\\
\chi'(r) &\simeq& \pm (1+2c_2 \beta_4 \phi_0^2)
\frac{\rho_0r_*^3}{6\beta_3 M_{\rm pl}^2 r^2} \,.
\ea
As in the case of Sec.~\ref{L3sec}, the behavior of the longitudinal
mode changes from $|\chi'| \propto r^{-1/2}$ to $|\chi'| \propto r^{-2}$
around $r=r_v$.

\begin{figure}
\begin{center}
\includegraphics[width=3.5in]{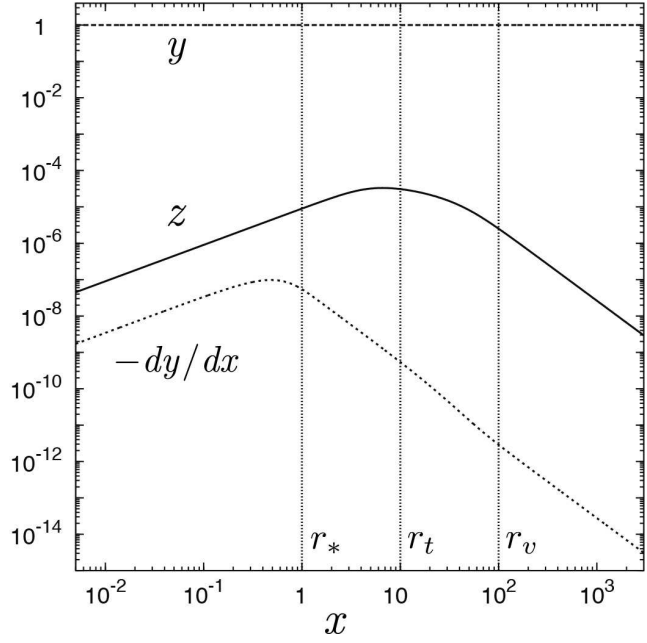}
\caption{The numerical solutions to $y=\phi/\phi_0$,
$-dy/dx$, and $z=\chi'/\phi_0$ as a function of $x=r/r_*$
for the theories with $s_{\beta_3}=1.0 \times 10^{-6}$ with $\beta_3>0$
and $\beta_4\phi_0^2=-1.6 \times 10^{-2}$, $c_2=1$,
and $\phi_0=1.0 \times 10^{-3}M_{\rm pl}$.
The matter profile is given by  $\rho_m=\rho_0 e^{-4r^2/r_*^2}$
with $\Phi_*=10^{-6}$.
The boundary conditions of $\Psi$, $\Phi$, $y$, and
$dy/dx$ are chosen to be consistent with Eqs.~(\ref{Sin}),
(\ref{phiap}), and (\ref{philinear}) at $x=10^{-3}$. The vertical lines stand for
the scales $r=r_*$, $r_t=10r_*$ and $r_v=100r_*$ respectively.
\label{fig2}}
\end{center}
\end{figure}

In Fig.~\ref{fig2} we plot an example of the field profile
derived by numerically solving the vector-field equations
of motion (\ref{F0full})-(\ref{F1full}) coupled with
the leading-order gravitational Eqs.~(\ref{GR1})-(\ref{GR2}).
The ratio $\beta_3/\beta_4$ is chosen to be negative
in this case, so the sign of $\chi'$ is positive.
Since $\varepsilon=s_{\beta_3}/(16\beta_4^2\phi_0^4)
\simeq 2.4 \times 10^{-4}$, the transition distance $r_t$
corresponds to $r_t \simeq 10 r_*$.
As estimated from Eqs.~(\ref{chiL4in}) and (\ref{chiinter}),
the numerical simulation of Fig.~\ref{fig2} shows that
$\chi'$ linearly grows in $r$ up to the distance
$r_t \simeq 10 r_*$. We also find that the longitudinal
mode behaves as $\chi'(r) \propto r^{-1/2}$ for
$r_t<r<r_v \simeq 100 r_*$ and
$\chi'(r) \propto r^{-2}$ for $r>r_v$.

As seen in Fig.~\ref{fig2}, the derivative of $\phi(r)$
has the dependence
$-\phi'(r) \propto r$ for $r \lesssim r_*$ and
$-\phi'(r) \propto r^{-2}$ for $r \gtrsim r_*$.
Since $|r\phi'(r)|$ is at most of the order of $\phi_0$ for
the whole distance range of interest,
the field $\phi$ stays nearly constant around $\phi_0$.
Thus, our numerical results are fully consistent with
the analytic solutions of $\chi'(r)$ and $\phi(r)$.

\subsection{Corrections to gravitational potentials}

Let us proceed to the calculations of corrections to
gravitational potentials $\Psi$ and $\Phi$ induced by the
vector field. We shall study the two regimes:
(i) $r_*<r<r_t$ and (ii) $r>r_t$, separately.

\subsubsection{$r_*<r<r_t$}

At this distance, the leading-order vector field solutions
are given by $\phi \simeq \phi_0$,
$\phi' \simeq -\phi_0 \Phi_* r_*/(3r^2)$,
and $\chi' \simeq -\beta_3 r/(8\beta_4)$,
where $\Phi_*=\rho_0r_*^2/M_{\rm pl}^2$.
In Eqs.~(\ref{GR1}) and (\ref{GR2}) we expand the
gravitational potentials $\Psi$, $\Phi$, and their derivatives
up to linear order.
The correction terms in Eqs.~(\ref{GR1d}) and (\ref{GR2d})
are approximately given by
\ba
\Delta_{\Phi}
&\simeq& -\frac{2\beta_4 \phi_0^4 \Phi_* \varepsilon
(\Phi_* \varepsilon x^2+3)}{3r_*^2}
-\frac{\phi_0^2\Phi_*^2}{18r_*^2x^4}
\,,\label{Del1}\\
\Delta_{\Psi}
&\simeq& -\frac{2\beta_4\phi_0^4\Phi_*
(5\Phi_* \varepsilon^2x^5+3\varepsilon x^3-3)}{9r_*^2x^3}
+\frac{\phi_0^2\Phi_*^2}{18r_*^2x^4}
\,,\label{Del2}
\ea
where we have employed the approximation (\ref{b4con})
and used the parameter $\varepsilon$ given by Eq.~(\ref{b4con2})
with $\rho_0=\Phi_*M_{\rm pl}^2/r_*^2$ and $x=r/r_*$.
Integrating Eqs.~(\ref{GR1d}) and (\ref{GR2d}) with these
corrections, the resulting gravitational potentials are
\ba
\Phi (r)
&\simeq& \frac{\Phi_* r_*}{6r}
\left[ 1-\frac{2\beta_4\phi_0^4\varepsilon x^3
(\Phi_*\varepsilon x^2+5)}{5\Mpl^2}
+\frac{\phi_0^2\Phi_*}{6\Mpl^2 x}
 \right]\,,\\
\Psi (r)
&\simeq&-\frac{\Phi_* r_*}{6r}
\left[ 1+\frac{2\beta_4\phi_0^4
(7\Phi_*\varepsilon^2x^5+15\varepsilon x^3+15)}
{15\Mpl^2}+\frac{\phi_0^2\Phi_*}{6\Mpl^2 x}
 \right]\,.
\ea
Provided that the corrections to
the leading-order gravitational potentials are small, the
post-Newtonian parameter $\gamma=-\Phi/\Psi$ reads
\be
\gamma \simeq 1-\frac{2\beta_4\phi_0^4(2\Phi_*\varepsilon^2x^5
+6\varepsilon x^3+3)}{3M_{\rm pl}^2}\,.
\label{gamL4}
\ee

At $r=r_*$ the first two terms inside the bracket of Eq.~(\ref{gamL4})
is sub-dominant to the last term, so Eq.~(\ref{gamL4}) reduces to
\be
\gamma \simeq
1-\frac{2\beta_4\phi_0^4}{M_{\rm pl}^2}\,.
\label{gamrstar}
\ee
The parameter $|\gamma-1|$ increases for larger $r$ and
it reaches the maximum value at $r=r_t$, i.e.,
\be
\gamma \simeq 1-\frac{3\beta_4 \phi_0^4}{M_{\rm pl}^2}\,.
\ee
Under the condition (\ref{b4con0}), the deviation of $\gamma$
from 1 is small. {}From the local gravity bound
$|\gamma-1|<2.3 \times 10^{-5}$, we obtain
\be
|\beta_4| \phi_0^4 <8 \times 10^{-6}M_{\rm pl}^2\,.
\label{b4con3}
\ee
This shows that, as long as the non-zero coupling
$\beta_3$ obeys Eq.~(\ref{b4con2}), the local
gravity constraint is satisfied under
the condition (\ref{b4con3}) for the distance $r<r_t$.

\subsubsection{$r>r_t$}

For $r$ larger than $r_t$, we only need to study the behavior
of $\Psi$ and $\Phi$ in the regime $r_t<r<r_v$ (because
$|\gamma-1|$ decreases for $r>r_v$ as we discussed
in Sec.~\ref{L3sec}).
The leading-order field solutions for $r_t<r<r_v$
are given by $\phi \simeq \phi_0$,
$\phi' \simeq -\phi_0 \Phi_* r_*/(3r^2)$
with the two branches of $\chi'$, i.e.,
$\chi' \simeq -\sqrt{\phi_0^2\Phi_*r_*/(12r)}$
for $\beta_3/\beta_4>0$ and
$\chi' \simeq \sqrt{\phi_0^2\Phi_*r_*/(12r)}$
for $\beta_3/\beta_4<0$. By using the relation
$\beta_3=\pm8\beta_4\phi_0\sqrt{\Phi_*\varepsilon}/(\sqrt{3}r_*)$,
the correction terms in Eqs.~(\ref{GR1d}) and (\ref{GR2d}) can be
expressed independently of the sign of $\beta_3/\beta_4$, as
\ba
\Delta_{\Phi} &\simeq&
-\frac{2\beta_4\phi_0^4\Phi_*\sqrt{\varepsilon}}{r_*^2x^{3/2}}
-\frac{\phi_0^2\Phi_*^2}{18r_*^2x^4}
\,,\label{DelPhil}\\
\Delta_{\Psi} &\simeq&
-\frac{\beta_4\phi_0^4\Phi_*
(5\Phi_*^2\sqrt{\varepsilon}-27\sqrt{x})}{54r_*^2x^{7/2}}
+\frac{\phi_0^2\Phi_*^2}{18r_*^2x^4}
\,,
\ea
where we used the approximation (\ref{b4con}).
The integrated solutions to gravitational potentials are
given by
\ba
\Phi (r)
&\simeq& \frac{\Phi_* r_*}{6r}
\left[ 1-\frac{4\beta_4\phi_0^4x^{3/2}\sqrt{\varepsilon}}{\Mpl^2}
+\frac{\phi_0^2\Phi_*}{6\Mpl^2x}
 \right]\,,\\
\Psi (r)
&\simeq& -\frac{\Phi_* r_*}{6r}
\left[ 1+\frac{\beta_4\phi_0^4(16x^{3/2}\sqrt{\varepsilon}+3)}{2\Mpl^2}
+\frac{\phi_0^2\Phi_*}{6\Mpl^2x}
\right]\,.
\label{Psil}
\ea
As long as the corrections to $\Phi_{\rm GR}$ and
$\Psi_{\rm GR}$ remain small, the post-Newtonian
parameter can be estimated as
\be
\gamma \simeq 1-\frac{3\beta_4\phi_0^4
(8x^{3/2}\sqrt{\varepsilon}+1)}
{2M_{\rm pl}^2}\,.
\ee

At $r=r_t$ this reduces to $\gamma-1 \simeq
-15\beta_4\phi_0^4/(2M_{\rm pl}^2)$, so the bound
$|\gamma-1|<2.3 \times 10^{-5}$ translates to
$|\beta_4|\phi_0^4<3 \times 10^{-6}$.
Taking into account Eq.~(\ref{b4con3}),
local gravity constraints can be satisfied for
\be
|\beta_4|\phi_0^4 \lesssim 10^{-6}M_{\rm pl}^2\,.
\label{b4con4}
\ee

At $r=r_v$ it follows that
\be
\gamma \simeq 1-3\frac{\phi_0^2}{M_{\rm pl}^2}\,.
\label{gamf}
\ee
Hence the resulting experimental bound is the same
as Eq.~(\ref{phi0}), i.e.,
\be
\phi_0 \lesssim 3 \times 10^{-3}M_{\rm pl}\,.
\label{phi0d}
\ee
If $r_v$ is far outside the solar-system scale,
we do not need to impose the condition (\ref{phi0d}).

\vspace{0.3cm}

In summary, under the conditions
(\ref{b4con4}) and (\ref{phi0d}) with $\beta_3$ in the range
(\ref{b4con2}), the deviation from GR is sufficiently small such that the
model is consistent with local gravity experiments.
When $\beta_3 \to 0$, it follows that $r_t$ goes to
infinity and that $\chi'$ vanishes for both $r<r_*$
and $r>r_*$. In the limit $\beta_3 \to 0$
(i.e., $\varepsilon \to 0$),
Eq.~(\ref{gamL4}) reduces to
$\gamma \simeq 1-2\beta_4\phi_0^4/M_{\rm pl}^2$,
so the local gravity bound is satisfied for
$|\beta_4|\phi_0^4 \lesssim 10^{-5}M_{\rm pl}^2$.
In this case, the deviation of $\gamma$ from 1
is directly related with the existence of the
$\beta_4X^2$ term in $G_4$.

\section{Conclusions}
\label{consec}

 In this paper, we have studied the screening mechanism of the fifth force
in a generalized class of Proca theories.
The breaking of $U(1)$ gauge invariance for an Abelian vector field
gives rise to non-trivial derivative self-interactions described by
the Lagrangians (\ref{L3})-(\ref{L5}), in addition to the Lagrangian
${\cal L}_2$ associated
with the mass term.
The equations of motion in these generalized Proca theories are
of second order without Ostrogradski instabilities,
while the number of propagating DOF remains
three (two transverse and one longitudinal)
as in the original Proca theory.

In the presence of a matter source, we derived the equations
of motion up to the Lagrangian ${\cal L}_4$ for a general
curved space-time and then applied them to a spherically
symmetric and static background described by
the line element (\ref{line}). First, we showed that
the transverse components of the spatial vector field $A^i$
vanish identically to satisfy the compatibility with
the spherically symmetric background and the regularity
of solutions at the origin.
Thus, we focused on the propagation of the
longitudinal scalar component of $A^i$ with
$A^{\mu}$ of the form (\ref{Amu}).

The leading-order gravitational interaction
in the vector-field equations should come from
the gravitational potentials $\Psi_{\rm GR}$ and
$\Phi_{\rm GR}$, whose interior and exterior solutions
around a compact body
($\rho_m \simeq \rho_0$ for $r<r_*$ and
$\rho_m \simeq 0$ for $r>r_*$) are given,
respectively, by Eqs.~(\ref{Sin}) and (\ref{Sout}).
After substituting these solutions into the vector
equations of motion under the weak-gravity approximation
($\Phi_*=\rho_0r_*^2/M_{\rm pl}^2 \ll 1$),
it is possible to derive analytic solutions of the
vector field $A^{\mu}$ (the temporal component $\phi$ and
the transverse mode $\chi'$) for a given Lagrangian.

In Sec.~\ref{L3sec} we obtained analytic vector field profiles and
corrections to the leading-order gravitational potentials
$\Phi_{\rm GR}$ and $\Psi_{\rm GR}$
in the presence of the vector Galileon
Lagrangian ${\cal L}_3=\beta_3 X\nabla_{\mu}A^{\mu}$
by assuming that the temporal component $\phi$ is of the form (\ref{phiap}).
Provided that the parameter $s_{\beta_3}=3(\beta_3\phi_0M_{\rm pl})^2/(4\rho_0)$
is larger than the order of 1, derivative self-interactions lead
to the suppression of the longitudinal mode $\chi'(r)$.
The fifth force can be screened in such a way that
the model is compatible with solar-system constraints of gravity.
For $s_{\beta_3} \ll 1$ the screening occurs partially at the
distance larger than $r_V$ given by Eq.~(\ref{rV}),
in which case the solar-system experiments lead to the bound
$\phi_0 \lesssim 3 \times 10^{-3}M_{\rm pl}$.
As shown in Fig.~\ref{fig1}, we have numerically confirmed that
our analytic solutions of the vector field are sufficiently trustable
even for the continuous density profile like Eq.~(\ref{denprofile}).

In Sec.~\ref{L4sec} we studied the vector Galileon theories up to
the Lagrangian ${\cal L}_4$ which contains a derivative self-coupling term
$\beta_4 X^2$ in the function $G_4$.
When $\beta_3=0$, we showed the existence of the solution where
$\chi'(r)$ vanishes everywhere.
If the Lagrangian ${\cal L}_3$ is present in addition to ${\cal L}_4$
and the former is subdominant to the latter, we obtained the solution
$\chi'(r)=-\beta_3r/(8\beta_4)$ for
$r\lesssim r_t=r_*/(4\varepsilon)^{1/3}$, where $\varepsilon$ is given by
Eq.~(\ref{b4con2}).
For $r>r_t$ the effect of the coupling $\beta_3$ manifests itself
in the longitudinal mode,
such that its amplitude decreases as $|\chi'(r)| \propto r^{-1/2}$ for
$r_t<r<r_v=r_t/(4 \beta_4^2 \phi_0^4)^{1/3}$ and
$|\chi'(r)| \propto r^{-2}$ for $r>r_v$ (see Fig.~\ref{fig2}).
The solar-system constraint at $r=r_t$ provides
a mild bound $|\beta_4|\phi_0^4 \lesssim 10^{-6}M_{\rm pl}^2$.
If $r_v$ is within the solar-system scale, we also obtain
the bound $\phi_0 \lesssim 3 \times 10^{-3}M_{\rm pl}$
from the estimation (\ref{gamf}) of the post-Newtonian parameter.

We have thus shown that the screening mechanism of the longitudinal
scalar for the vector field is at work in the presence of cubic and
quartic derivative self-interactions. It will be of interest to study whether
the similar mechanism holds or not with the Lagrangian ${\cal L}_5$.
We leave this analysis for a future work.

\section*{Acknowledgements}
We thank Xian Gao, Kazuya Koyama, and Gianmassimo Tasinato
for valuable discussions.
LH acknowledges financial
support from Dr.~Max R\"ossler, the Walter Haefner Foundation and
the ETH Zurich Foundation.
RK is supported by the Grant-in-Aid for Research Activity Start-up
of the JSPS No.\,15H06635.
ST is supported by the Grant-in-Aid for Scientific Research Fund of
the JSPS No.\,24540286, MEXT KAKENHI Grant-in-Aid for Scientific Research
on Innovative Areas ``Cosmic Acceleration'' (No.\,15H05890),
and the cooperation program between Tokyo
University of Science and CSIC. YZ and GZ are supported by
the Strategic Priority Research Program
``The Emergence of Cosmological Structures'' of the Chinese
Academy of Sciences, Grant No. XDB09000000.

\section*{Appendix Expressions for the coefficients $\mathcal{C}_i$}

The coefficients of the gravitational
Eqs.~(\ref{eom00})-(\ref{eom22}) are given by
\ba
&&
\C_1=4\Xp [1-2c_2(G_{4,X}+2\Xp G_{4,XX})]\,,
\notag\\
&&
\C_2=4\chi' \Xp G_{3,X}
+2e^{2\Psi}\phi\phi'[1-2c_2(G_{4,X}+2\Xp G_{4,XX})]\,,
\notag\\
&&
\C_3=-32\Xp\Xc G_{4,XX}\,,\quad
\C_4=-2\chi' (\Xp+\Xc) G_{3,X}\,,
\notag\\
&&
\C_5=-4[G_4-2(\Xp+2\Xc)G_{4,X}-4\Xc (\Xp+\Xc) G_{4,XX}]\,,
\notag\\
&&
\C_6=-e^{2\Phi}(G_2-2\Xp G_{2,X})+[e^{2\Psi}\phi\phi'\chi'
+2\chi''(\Xp+\Xc)] G_{3,X}+\frac{1}{2}e^{2\Psi}\phi'^2
[1-2c_2(G_{4,X}+2\Xp G_{4,XX})]\,,
\notag\\
&&\C_7=4\chi'\Xp G_{3,X}+4e^{-2\Phi}\chi'\chi'' G_{4,X}
+8[e^{-2\Phi}\chi'\chi''(\Xp+\Xc)-e^{2\Psi}\phi\phi'\Xc]G_{4,XX}\,,
\notag\\
&&
\C_8=2(1-e^{2\Phi})G_4-4[\Xc+(1-e^{2\Phi})\Xp]G_{4,X}-8\Xp\Xc G_{4,XX}\,,
\notag\\
&&
\C_9=4\Xp[1-2c_2(G_{4,X}+2\Xc G_{4,XX})]\,,
\notag\\
&&
\C_{10}=2\chi'(\Xc-\Xp)G_{3,X}+2e^{2\Psi}\phi\phi'
[1-2c_2(G_{4,X}+2\Xc G_{4,XX})]\,,
\notag\\
&&
\C_{11}=4[G_4+2(\Xp-2\Xc)G_{4,X}+4\Xc(\Xp-\Xc)G_{4,XX}]\,,
\notag\\
&&
\C_{12}=-e^{2\Phi}(G_2-2\Xc G_{2,X})-e^{2\Psi}\phi\phi'\chi'G_{3,X}
+\frac{1}{2}e^{2\Psi}\phi'^2[1-2c_2(G_{4,X}+2\Xc G_{4,XX})]\,,
\notag\\
&&
\C_{13}=4\chi'\Xc G_{3,X}+4e^{2\Psi}\phi\phi'(G_{4,X}+2\Xc G_{4,XX})\,,
\notag\\
&&
\C_{14}=2(1-e^{2\Phi})G_4-4\Xc (2-e^{2\Phi})G_{4,X}-8\Xc^2 G_{4,XX}\,,
\notag\\
&&
\C_{15}=2[G_4+2(\Xp-\Xc)G_{4,X}]\,,
\notag\\
&&
\C_{16}=2[G_4+2\{2(c_2+2)\Xp-\Xc\}G_{4,X}+4\Xp(\Xp-\Xc)G_{4,XX}-2\Xp]\,,
\notag\\
&&
\C_{17}=-2[G_4+2(\Xp-2\Xc)G_{4,X}+4\Xc(\Xp-\Xc)G_{4,XX}]\,,
\notag\\
&&
\C_{18}=2\chi'\Xp G_{3,X}-2e^{2\Psi}\phi\phi'[1-2(c_2+3)G_{4,X}
\notag\\
&&
\hspace{1cm}+2(\Xc-2\Xp) G_{4,XX}]
+2e^{-2\Phi}\chi'\chi''[G_{4,X}+2(\Xc-\Xp)G_{4,XX}]\,,
\notag\\
&&
\C_{19}=-2[G_4-4\Xc (G_{4,X}+\Xc G_{4,XX})]\,,\notag\\
&&
\C_{20}=-e^{2\Phi}G_2+(2\chi''\Xc+e^{2\Psi}\phi\phi'\chi') G_{3,X}
-\frac{1}{2}e^{2\Psi}\phi'^2[1-2(c_2+2)G_{4,X}-8\Xp G_{4,XX}]
\notag\\
&&\hspace{1cm}
+2e^{2\Psi}\phi\phi''G_{4,X}-2e^{2\Psi-2\Phi} \phi\phi'\chi'\chi''G_{4,XX}\,,
\notag\\
&&
\C_{21}=2e^{2\Psi}\phi\phi'(G_{4,X}-2\Xc G_{4,XX})
+2e^{-2\Phi}\chi'\chi''(G_{4,X}+2\Xc G_{4,XX})\,.
\ea
%



\begin{thebibliography}{99}

\bibitem{CST}
E.~J.~Copeland, M.~Sami and S.~Tsujikawa,
Int.\ J.\ Mod.\ Phys.\ D {\bf 15}, 1753 (2006).

\bibitem{Sil}
A.~Silvestri and M.~Trodden,
Rept.\ Prog.\ Phys.\  {\bf 72}, 096901 (2009).

\bibitem{Tsuji10}
S.~Tsujikawa,
Lect.\ Notes Phys.\  {\bf 800}, 99 (2010).

\bibitem{Pedro}
T.~Clifton, P.~G.~Ferreira, A.~Padilla and C.~Skordis,
Phys.\ Rept.\  {\bf 513}, 1 (2012).

\bibitem{Joyce}
A.~Joyce, B.~Jain, J.~Khoury and M.~Trodden,
Phys.\ Rept.\  {\bf 568}, 1 (2015).

\bibitem{Ostro}
M.~V.~Ostrogradski, Mem. Acad. St. Petersbourg VI 4,
{\bf 385} (1850).

\bibitem{Horndeski}
G.~W.~Horndeski,
Int.\ J.\ Theor.\ Phys.\ 10, 363-384 (1974).

\bibitem{Nicolis}
A.~Nicolis, R.~Rattazzi and E.~Trincherini,
Phys.\ Rev.\ D {\bf 79}, 064036 (2009).

\bibitem{Defa1}
C.~Deffayet, G.~Esposito-Farese and A.~Vikman,
Phys.\ Rev.\ D {\bf 79}, 084003 (2009).

\bibitem{Defa2}
C.~Deffayet, X.~Gao, D.~A.~Steer and G.~Zahariade,
Phys.\ Rev.\ D {\bf 84}, 064039 (2011).

\bibitem{KYY}
T.~Kobayashi, M.~Yamaguchi and J.~'i.~Yokoyama,
Prog.\ Theor.\ Phys.\  {\bf 126}, 511 (2011).

\bibitem{Gao}
X.~Gao,
JCAP {\bf 1110}, 021 (2011).

\bibitem{GS}
X.~Gao and D.~A.~Steer,
JCAP {\bf 1112}, 019 (2011).

\bibitem{CdRLH}
C.~de Rham and L.~Heisenberg,
Phys.\ Rev.\ D {\bf 84}, 043503 (2011).

\bibitem{Char}
C.~Charmousis, E.~J.~Copeland, A.~Padilla and P.~M.~Saffin,
Phys.\ Rev.\ Lett.\  {\bf 108}, 051101 (2012).

\bibitem{LHRKKY}
L.~Heisenberg, R.~Kimura and K.~Yamamoto,
Phys.\ Rev.\ D {\bf 89}, 103008 (2014).

\bibitem{Horndeski2}
G.~W.~Horndeski,
J.\ Math.\ Phys.\  {\bf 17}, 1980 (1976).

\bibitem{Barrow}
J.~D.~Barrow, M.~Thorsrud and K.~Yamamoto,
JHEP {\bf 1302}, 146 (2013).

\bibitem{Jimenez}
J.~B.~Jimenez, R.~Durrer, L.~Heisenberg and M.~Thorsrud,
JCAP {\bf 1310}, 064 (2013).

\bibitem{Deser}
C.~Deffayet, S.~Deser and G.~Esposito-Farese,
Phys.\ Rev.\ D {\bf 82}, 061501 (2010).

\bibitem{TKK}
G.~Tasinato, K.~Koyama and N.~Khosravi,
JCAP {\bf 1311}, 037 (2013).

\bibitem{Fleury}
P.~Fleury, J.~P.~B.~Almeida, C.~Pitrou and J.~P.~Uzan,
JCAP {\bf 1411}, 043 (2014).

\bibitem{Mukohyama}
C.~Deffayet, A.~E.~Gumrukcuoglu, S.~Mukohyama and Y.~Wang,
JHEP {\bf 1404}, 082 (2014).

\bibitem{Mukohyama2}
C.~Deffayet, S.~Mukohyama and V.~Sivanesan,
arXiv:1601.01287 [hep-th].

\bibitem{Heisenberg}
L.~Heisenberg,
JCAP {\bf 1405}, 015 (2014).

\bibitem{Tasinato1}
G.~Tasinato,
JHEP {\bf 1404}, 067 (2014).

\bibitem{Tasinato2}
G.~Tasinato,
Class.\ Quant.\ Grav.\  {\bf 31}, 225004 (2014).

\bibitem{Hull}
M.~Hull, K.~Koyama and G.~Tasinato,
JHEP {\bf 1503}, 154 (2015).

\bibitem{Li}
W.~Li,
arXiv:1508.03247 [gr-qc].

\bibitem{Hull2}
M.~Hull, K.~Koyama and G.~Tasinato,
arXiv:1510.07029 [hep-th].

\bibitem{Peter}
E.~Allys, P.~Peter and Y.~Rodriguez,
arXiv:1511.03101 [hep-th].

\bibitem{JBJTSK}
J.~Beltran Jimenez and T.~S.~Koivisto,
Class.\ Quant.\ Grav.\  {\bf 31}, 135002 (2014).

\bibitem{Vainshtein}
A.~I.~Vainshtein,
Phys.\ Lett.\ B \textbf{39}, 393 (1972).

\bibitem{Burrage}
C.~Burrage and D.~Seery,
JCAP {\bf 1008}, 011 (2010).

\bibitem{DKT}
A.~De Felice, R.~Kase and S.~Tsujikawa,
Phys.\ Rev.\ D {\bf 85}, 044059 (2012).

\bibitem{KKY}
R.~Kimura, T.~Kobayashi and K.~Yamamoto,
Phys.\ Rev.\ D {\bf 85}, 024023 (2012).

\bibitem{Kase13}
R.~Kase and S.~Tsujikawa,
JCAP {\bf 1308}, 054 (2013)
[arXiv:1306.6401 [gr-qc]].

\bibitem{Stu}
E.~Stueckelberg, Helv. Phys. Acta, {\bf 11}, 225 (1938).

\bibitem{Will}
C.~M.~Will,
Living Rev.\ Rel.\  {\bf 9}, 3 (2006).


\end{thebibliography}
\end{document}